\documentclass[fleqn,usenatbib]{mnras}

\usepackage{newtxtext,newtxmath}

\usepackage[T1]{fontenc}

\DeclareRobustCommand{\VAN}[3]{#2}
\let\VANthebibliography\thebibliography
\def\thebibliography{\DeclareRobustCommand{\VAN}[3]{##3}\VANthebibliography}

\usepackage{graphicx}	
\usepackage{amsmath}	
\usepackage{natbib}
\usepackage{microtype}
\usepackage{xcolor}
\usepackage{multirow,tabularx}
\usepackage[toc,page]{appendix}

\newcommand{\bham}{School of Physics and Astronomy, University of Birmingham, Birmingham B15 2TT, UK}
\newcommand{\igw}{Institute for Gravitational Wave Astronomy, University of Birmingham, Birmingham B15 2TT, UK}

\newcommand{\berkeley}{Astronomy Department and Theoretical Astrophysics Center, University of California, Berkeley, Berkeley, CA 94720, USA}

\def\M{M$_\odot$}
\def\Mtov{$M_{\rm TOV}$}
\def\Rns{$R_{1.4}$}

\title[Kilonova light curves from binary parameters]{
Tight multi-messenger constraints on the neutron star equation of state from GW170817 and a forward model for kilonova light curve synthesis}

\author[M.~Nicholl et al]{{Matt Nicholl$^{1,2}$}\thanks{Contact e-mail: \href{mailto:m.nicholl.1.bham.ac.uk}{m.nicholl.1.bham.ac.uk}}, Ben Margalit$^{3}$\thanks{NASA Einstein Fellow}, Patricia Schmidt$^{1,2}$, Graham P.~Smith$^{1}$, Evan J.~Ridley$^{1,2}$ \& James Nuttall$^{1}$
\\
$^{1}$\bham\\
$^{2}$\igw\\
$^{3}$\berkeley
}

\date{}

\pubyear{2021}

\begin{document}
\label{firstpage}
\pagerange{\pageref{firstpage}--\pageref{lastpage}}
\maketitle


\begin{abstract}
We present a rapid analytic framework for predicting kilonova light curves following neutron star (NS) mergers, where the main input parameters are binary-based properties measurable by gravitational wave detectors (chirp mass and mass ratio, orbital inclination) and properties dependent on the nuclear equation of state (tidal deformability, maximum NS mass). This enables synthesis of a kilonova sample for any NS source population, or determination of the observing depth needed to detect a live kilonova given gravitational wave source parameters in low latency. We validate this code, implemented in the public \textsc{mosfit} package, by fitting it to GW170817. A Bayes factor analysis overwhelmingly ($B>10^{10}$) favours the inclusion of an additional luminosity source in addition to lanthanide-poor dynamical ejecta during the first day. This is well fit by a shock-heated cocoon model, though differences in the ejecta structure, opacity or nuclear heating rate cannot be ruled out as alternatives. The emission thereafter is dominated by a lanthanide-rich viscous wind. We find the mass ratio of the binary is $q=0.92\pm0.07$ (90\% credible interval). We place tight constraints on the maximum stable NS mass, \Mtov\,$=2.17^{+0.08}_{-0.11}$\,\M. For a uniform prior in tidal deformability, the radius of a 1.4\,\M\ NS is \Rns\,$\sim 10.7$\,km. Re-weighting with a prior based on equations of state that support our credible range in \Mtov, we derive a final measurement \Rns\,$=11.06^{+1.01}_{-0.98}$\,km. Applying our code to the second gravitationally-detected neutron star merger, GW190425, we estimate that an associated kilonova would have been fainter (by $\sim0.7$\,mag at one day post-merger) and declined faster than GW170817, underlining the importance of tuning follow-up strategies individually for each GW-detected NS merger.
\end{abstract}




\section{Introduction}

The first binary neutron star (NS) merger detected through its gravitational wave emission, GW170817 \citep{Abbott2017a}, was accompanied by both a short gamma-ray burst \citep[GRB;][]{Goldstein2017,Savchenko2017} and an optical counterpart \citep{Coulter2017,SoaresSantos2017,Valenti2017,Arcavi2017,Tanvir2017,Lipunov2017}. Thermal transients from these mergers had long been predicted in the form of a kilonova \citep{Li1998,Rosswog1999,Metzger2010}: a non-relativistic outflow heated by the decays of heavy nuclei formed through rapid neutron captures \citep[the so-called `r-process';][]{Lattimer1974,Eichler1989,Davies1994,Freiburghaus1999}. Extensive optical and infrared follow-up of GW170817 by many groups showed photometric and spectroscopic properties that were remarkably consistent with the expectations for an r-process kilonova \citep{Andreoni2017,Arcavi2017,Chornock2017,Cowperthwaite2017,Diaz2017,Drout2017,Evans2017,Hu2017,Kasliwal2017,McCully2017,Nicholl2017,Pian2017,Pozanenko2018,Shappee2017,Smartt2017,Tanvir2017,Troja2017,Utsumi2017,Valenti2017}.

Studying these sources is crucial for understanding cosmic chemical evolution. The total ejected mass powering the GW170817 kilonova was inferred to be $\sim 0.05$\,\M, with at least two components of different compositions and hence opacities \citep[e.g.][]{Villar2017,Kasen2017,Margutti2020}: one containing light r-process elements (atomic mass number $A\lesssim140$), and one containing the heavier lanthanides, which absorb optical photons in the forest of transitions in their open $f$-shells \citep{Kasen2013,Barnes2013}.
The large total mass and broad range of mass numbers covering all three r-process abundance peaks \citep{Burbidge1957} suggested that NS mergers may be the dominant production site of heavy nuclei in the Universe.
 
Another central goal of multi-messenger astronomy is to determine the NS equation of state (EoS), i.e.~the relation between pressure and density. This provides unique tests of nuclear physics beyond the saturation density of nuclear matter \citep[e.g.][]{Lattimer2016}. Because the EoS determines the NS radius and maximum stable mass, measuring these quantities from astrophysical observation directly constrains the EoS. For example, a viable EoS must support the existence of the most massive known pulsars, with mass $\gtrsim 2$\,\M\ \citep{Demorest2010,Antoniadis2013,Cromartie2020}. NS mergers provide an excellent laboratory for EoS measurements. The degree of tidal deformation imprinted in the GW signal \citep{Flanagan2008,Damour2012} and the strength of shocks at the contact interface \citep{Bauswein2013,Hotokezaka2013} both depend on the radius, typically parameterised as \Rns, the radius of a 1.4\,\M\ NS. Similarly, the lifetime of the massive merger product until collapse to a black hole (BH) depends on the maximum stable NS mass, the Tolman-Oppenheimer-Volkoff mass \Mtov\ \citep{Shapiro1986,Margalit2017}.

So far, most of the EoS constraints for GW170817 have come from the GW data \citep[e.g.][]{Abbott2018,De2018,Raithel2018}. However, electromagnetic data have also provided important insight. \citet{Nicholl2017} pointed out that the inferred combination of large mass and high velocity in the low-lanthanide ejecta was predicted only in the case of relatively small radii \citep{Bauswein2013,Shibata2019}. More quantitatively, \citet{Margalit2017} used the 
observed electromagnetic signatures of GW170817 to infer that the merged remnant must have collapsed to a BH within a short timescale ($\sim 10$\,ms to $\sim 1$\,s),
pointing to \Mtov\,$\lesssim2.17$\,\M. \citet{Bauswein2017} related this lifetime to the radius, finding \Rns\,$>10.68$\,km. 

A major caveat in these analyses is our ability to associate a measured ejecta component with a specific physical origin. NS mergers can eject mass dynamically during the merger, through tidal stripping or shock heating, and after the merger, via winds from an accretion disk or long-lived remnant \citep[see review by][]{Metzger2019}. Each mechanism is expected to produce a different combination of ejecta mass, velocity and composition, and the relative importance of each mechanism depends on the properties of the system before merger \citep[e.g.][]{Bauswein2013,Hotokezaka2013,Sekiguchi2016,Radice2018,Shibata2019,Ciolfi2020}. 

Most model fits to the electromagnetic data from GW170817 have been specified in terms of the `post-merger' parameters: i.e.~the individual masses, velocities and compositions/opacities of some number of ejecta components. This introduces subjectivity into the origin of each component, and neglects physically-expected correlations between the properties of the different components. Specifying instead a forward model in terms of `pre-merger' properties -- binary masses and mass ratios, and the EoS -- would allow one to trace explicitly the ejection mechanism of each component, and exclude models that fit the data well but require unphysical combinations of ejecta parameters. Moreover, analysing the data in terms of pre-merger properties provides a more direct link between the electromagnetic and GW signals.

Progress in combining light curve models with GW inference was first made by \citet{Coughlin2019}, who fit the light curve of the GW170817 kilonova and related the implied ejecta masses to the predictions of numerical relativity simulations for different system mass ratios and combinations of \Rns\ and \Mtov. Including GW information in their priors, they measured \Rns\,$=11.3-13.5$\,km. However, their model was unable to match the early blue part of the light curve, possibly because a sufficiently high mass of low-lanthanide ejecta was not predicted by merger simulations at the GW-inferred binary mass of GW170817. This may suggest the need to look at alternative luminosity sources for the early emission, such as cooling of matter shock-heated by the GRB jet \citep{Kasliwal2017,Piro2018,Arcavi2018}. 
While our manuscript was in preparation, other efforts have built on these techniques. \citet{Dietrich2020} published an updated analysis, using a light curve model that reasonably reproduced the early flux without shock cooling, and found \Rns\,$=11.75\pm0.86$\,km. An even more recent analysis of GW170817 was published by \citet{Breschi2021}, who found \Rns\,$=12.2\pm0.5$\,km using a three-component kilonova model.

A better understanding of NS physics will come with larger samples of kilonovae, but finding these events has proven challenging.
Despite subsequent GW detections of binaries hosting possible NSs, such as GW190425 \citep{Abbott2020a} and GW190814 \citep{Abbott2020b}, and significant effort invested in counterpart searches \citep{Hosseinzadeh2019,Lundquist2019,Coughlin2019b,Gomez2019,Andreoni2020,Vieira2020,Engrave2020,Kasliwal2020,Gompertz2020,Paterson2020,Anand2021},
at the end of the third LIGO-Virgo observing run (O3), GW170817 remains the only NS binary detected in both electromagnetic and gravitational waves \citep[but see][for a possible optical counterpart to a BH binary merger]{Graham2020}. 
Analysis of short GRB afterglows has yielded several kilonova candidates \citep{Berger2013,Tanvir2013,Jin2016,Troja2018,Troja2019,Gompertz2018,Lamb2019,Fong2020} and constraining non-detections \citep{Rastinejad2021,OConnor2021}, while optical surveys are also looking for kilonovae in their data streams, though without an unambiguous detection to date \citep{McBrien2020,Andreoni2020b}. However, a step change in our ability to find kilonovae is anticipated, due to planned improvements in the sensitivity of GW detector networks \citep{ligo2020}, and the deepest ever time-domain sky survey: the Vera Rubin Observatory Legacy Survey of Space and Time \citep[LSST;][]{lsst2009}.

In this paper we provide a fast, analytic forward model for kilonova light curves that is completely specified by the binary configuration and EoS-dependent properties. Fitting such a model to observed kilonova data allows one to measure fundamental pre-merger properties even in the absence of a GW signal, or to use GW information when available as priors for multi-messenger inference. Forward modelling the light curves of kilonovae can increase the efficiency of GW-triggered optical searches, by predicting the electromagnetic light curve directly from the GW signal and providing the required observing depth to detect (or rule out) a counterpart at any time since merger \citep[see also][]{Salafia2017,Barbieri2020,Stachie2021}. In the context of all-sky surveys like LSST, such a model will allow for kilonova population synthesis \citep[see also][]{Kashyap2019,Mochkovitch2021} from a given NS source population, informing strategies for selecting candidate kilonovae from the data stream or allowing one to constrain the NS merger rate even given non-detections.

GW170817 serves as an ideal testing ground for such a model. By fitting the rich observed data set, we will show the power of using GW results as priors, constrain the physical origins of the different emission components, and present new measurements of the mass ratio of the progenitor system, the viewing angle, and the EoS-dependent quantities \Rns\ and \Mtov.


The structure of this paper is as follows. In section \ref{sec:model}, we describe the model setup and assumptions. Section \ref{sec:170817} shows its application to GW170817 and our main results. We briefly demonstrate the utility of our code for aiding future kilonova searches, using the example of GW190425, in section \ref{sec:predict}, before concluding in section \ref{sec:conc}. A set of kilonova model light curves produced using our code is available for download (see Data Availability section).

\begin{figure*}
    \centering
    \includegraphics[width=15cm]{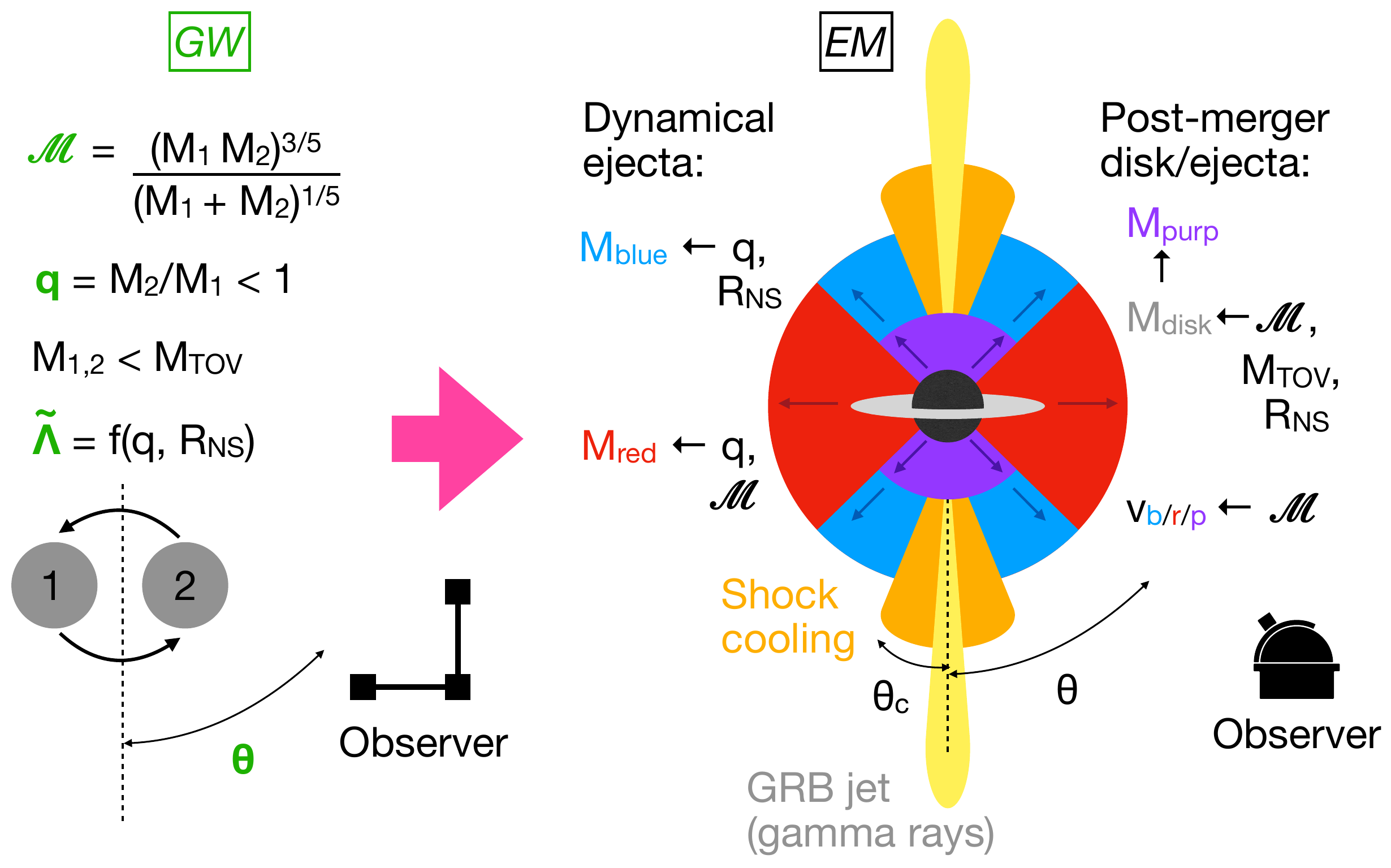}
    \caption{Schematic showing model geometry and relations between parameters. A full description of these parameters is given in section \ref{sec:model}. Left: Pre-merger properties of a NS binary. Bold, green font is used to indicate properties that can be measured from GW waveforms. Right: Ejecta properties probed by electromagnetic observations. Left arrows are used to mean `depends on', in order to highlight the most important parameter sensitivities. The merged remnant can be a long-lived or short-lived NS or a BH, depending on the ratio $(M_1+M_2) / M_{\rm TOV}$. The blue ejecta are from shocks at the collisional interface and have an opacity $\kappa_{\rm blue}\sim0.5$\,cm$^2$\,g$^{-1}$, the red ejecta are tidally stripped in the orbital plane and have $\kappa_{\rm red}\sim10$\,cm$^2$\,g$^{-1}$, and the purple ejecta are winds from an accretion disk and have an intermediate opacity that depends on the lifetime of the remnant. Low opacity ejecta may not be visible for large viewing angle $\theta$. The shock cooling emission arises from a cocoon of width $\theta_c$ heated at its interface with the GRB jet.}
    \label{fig:schem}
\end{figure*}

\section{Description of model}
\label{sec:model}

We have built our model using the Modular Open Source Fitter for Transients \citep[\textsc{mosfit};][]{Guillochon2018}. Modules to calculate a radioactive kilonova light curve for a given set of ejecta masses, velocities and opacities have already been implemented by \citet{Cowperthwaite2017} and \citet{Villar2017}. Therefore our main task here is to define modules that provide these masses, velocities and opacities for a given binary configuration. We also provide new modules to take into account the effects of observer viewing angle and shock cooling. We have made these new modules and the parameter files to implement them publicly available via GitHub\footnote{\url{https://github.com/guillochon/MOSFiT}}, as a new model called `\textsc{bns}'. In this section we will describe the main features. Figure \ref{fig:schem} provides a visual overview that summarises the geometry and the relations between key parameters. The connection between our modelling framework and the GW observables is discussed in detail in section \ref{sec:gw}.

\subsection{Dynamical ejecta}
\label{sec:dyn}

During the merger process, mass is ejected dynamically both by tidal forces in the orbital plane \citep{Rosswog1999,Sekiguchi2015} and by shocks at the contact interface between the two NSs \citep{Bauswein2013,Hotokezaka2013}. To estimate the mass of this material, we use a fit by \citet{Dietrich2017} to the ejecta masses of 172 numerical relativity simulations. These simulations, from \citet{Hotokezaka2013,Bauswein2013,Dietrich2015,Dietrich2017b,Lehner2016,Sekiguchi2016}, cover a range of binary masses, though all neglect the effect of NS spin. The catalog includes both grid-based and smoothed particle hydrodynamics simulations, and simulations with and without neutrino treatment. In total they cover 21 simplified or tabulated equations of state.

Their function takes the form
\begin{multline}
\label{eq:Mdyn}
    \frac{M_{\rm dyn}}{10^{-3} {\rm M}_\odot} = \Bigg[a\left(\frac{M_2}{M_1}\right)^{1/3} \left(\frac{1-2C_1}{C1}\right) + \\ b\left(\frac{M_2}{M_1}\right)^n + c\left(1-M_1/M_1^*\right) \Bigg] M_1^* + (1\leftrightarrow2) + d,
\end{multline}
where $M_{i}$ is the mass of the $i$th NS and 
\begin{equation}
\label{eq:C}
    C_i \equiv G M_{i} / R_{i} c^2
\end{equation}
is its compactness given a radius $R_i$. The term $1\leftrightarrow2$ represents exchanging the subscripts, and the best-fit values of the free parameters $a,b,c,d,n$ can be found in their study. Quantities denoted $M_i^*$ are the baryonic masses (differing from the gravitational mass by the binding energy), which we calculate for a given $M_i$ (in solar masses) following \citet{Gao2020}: $M_i^*=M_i+0.08M_i^2$. The simulations cover the ranges $1<M_i/M_{\odot}<2$ and $0.1<C_i<0.23$. The typical fractional uncertainty on $M_{\rm dyn}$ using this parameterisation was found by \citet{Dietrich2017} to be $\approx 72\%$. Similar expressions were provided for the equatorial and polar ejecta velocities, with smaller uncertainties of only $13-33\%$.


Inspection of equation \ref{eq:Mdyn} shows that the dynamical ejecta mass depends on terms of the form $M_i/M_j$. We adopt the convention that $M_1$ is the heavier NS (for consistency with GW analyses, e.g. \citealt{Abbott2017a,Abbott2018}), such that the mass ratio of the system is $q=M_2/M_1<1$. The other important sensitivity in equation \ref{eq:Mdyn} is to the compactness $C_{i}$, an EoS-dependent quantity. This will be discussed further in section \ref{sec:gw}.

The relationship between binary and ejecta masses for NS mergers was recently revisited by \citet{Kruger2020} and \citet{Nedora2020}, who used a different set of numerical data but found broadly consistent fits. \citet{Coughlin2019} provided a fitting formula in terms of $\log(M_{\rm dyn})$, which performed similarly well. However, we find this logarithmic form can lead to unrealistically large ejecta masses in some regions of parameter space, which is not optimal for population synthesis.

To better capture the multiple physical processes contributing to the dynamical ejecta, we divide it into two components. The tidal ejecta, concentrated in the equatorial plane, are expected to contain a very low electron fraction (the ratio of protons to total nucleons), $Y_e < 0.25$, allowing the complete r-process synthesis of very heavy nuclei. These ejecta have been termed `red' due to the high opacity of the resulting lanthanides at optical wavelengths \citep{Barnes2013}. The polar material has a higher $Y_e > 0.25$, due to the importance of shocks and neutrino heating at higher latitudes \citep[e.g.][]{Bauswein2013,Sekiguchi2016,Metzger2019}, and is correspondingly `blue' due to an incomplete r-process only up to mass number $A\sim 140$. \citet{Radice2018} find that the ratio of red to blue ejecta is largely insensitive to the total system mass or the EoS, and depends primarily on the mass ratio of the system.

\begin{figure}
    \centering
    \includegraphics[width=\columnwidth]{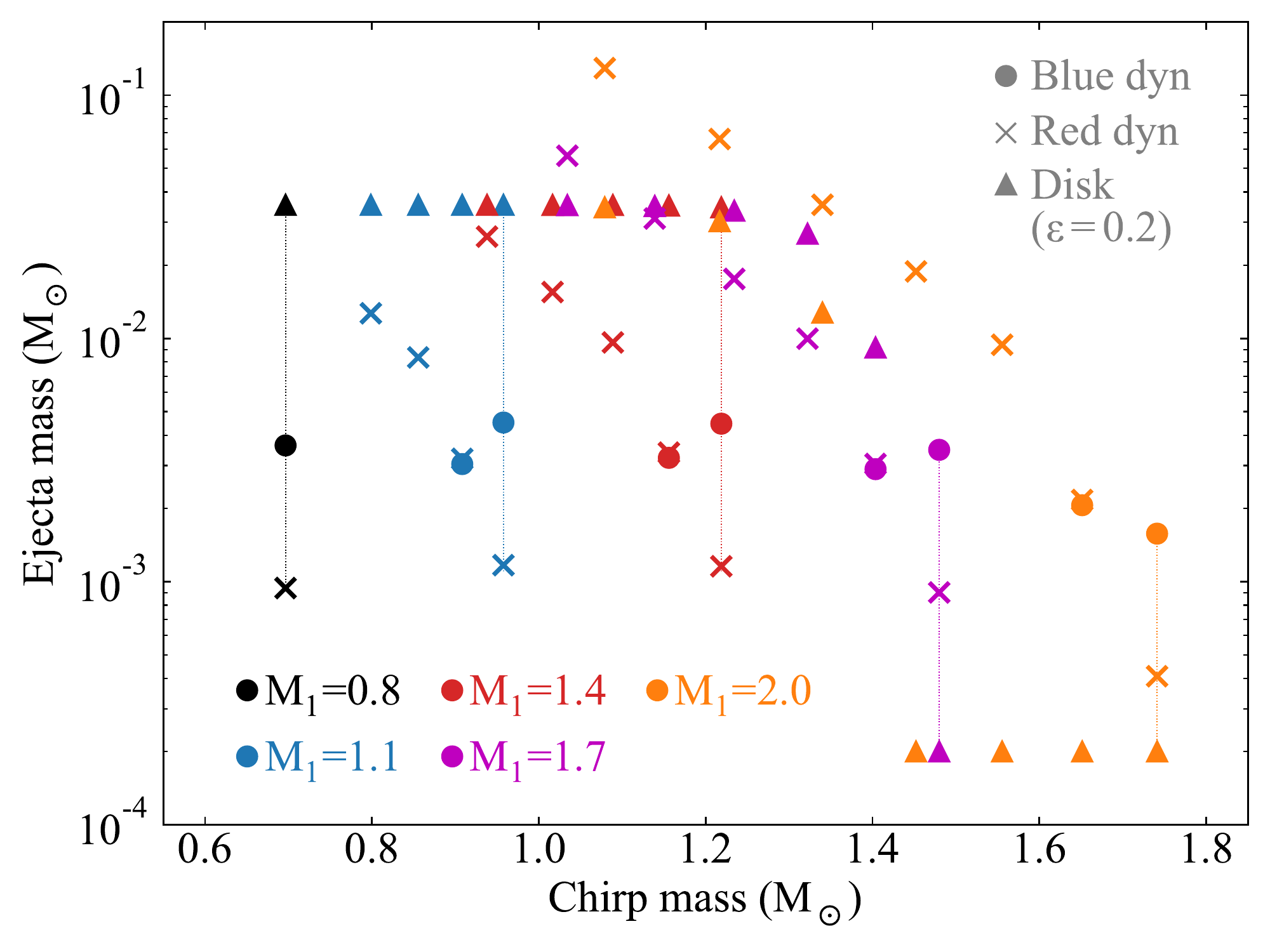}
    \includegraphics[width=\columnwidth]{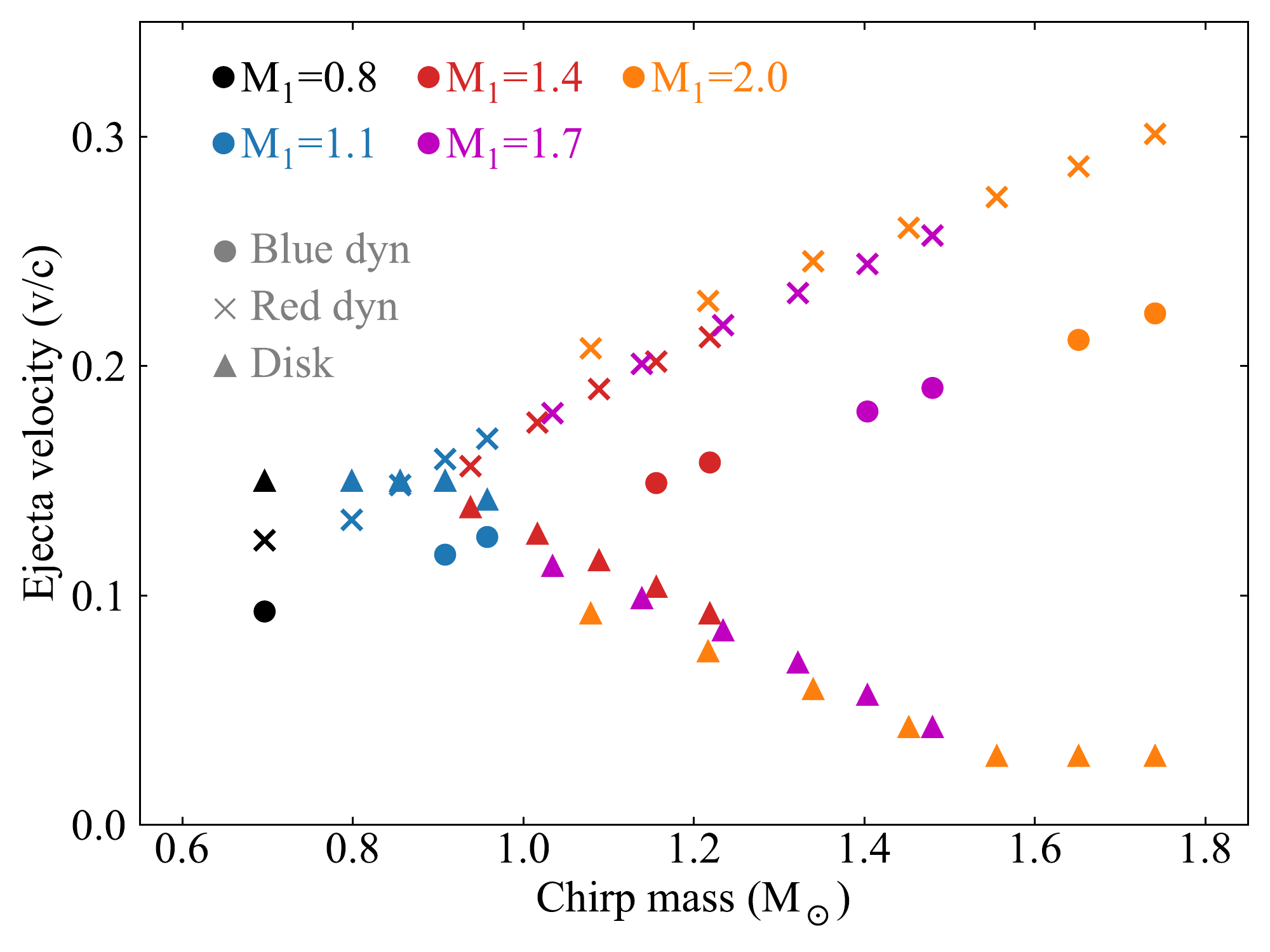}
    \caption{Top: Predicted ejecta mass in each component (blue and red dynamical, and purple disk ejecta) using relations from \citet{Dietrich2017} and \citet{Coughlin2019}, assuming $R_{1.4}=11$\,km, $M_{\rm TOV}=2.1$\,\M, and that 20\% of the remnant disk is ejected. Each colour corresponds to a fixed primary mass, $M_1$, with varying mass ratio, $0.4\le q\le1$. Vertical lines indicate $M_2=M_1$, with lower mass secondaries (less equal binaries) to the left. Bottom: Predictions for ejecta velocities. Disk ejecta velocities are estimated using results from \citet{Metzger2014}. }
    \label{fig:mass}
\end{figure}

To estimate the relative mass in the red and blue dynamical ejecta, we use simulations by \citet{Sekiguchi2016}. They provide $Y_e$ distributions in the dynamical ejecta as a function of $q$ for two equations of state, one stiff (large radius) and one soft (compact). We integrate these distributions (in their Figure 5) to determine $f_{\rm red}(q) \equiv M(Y_e<0.25)/M({\rm total})$ and $f_{\rm blue} \equiv 1-f_{\rm red}$. We find that $f_{\rm red}$ is insensitive to the EoS but is a steep function of $q$, ranging from 20\% for an equal-mass merger ($q=1$) to 76\% for $q=0.86$. For $q\lesssim 0.8$, $f_{\rm red}\approx 1$, because the less massive NS is tidally disrupted prior to contact, preventing strong shocks \citep[e.g.][]{Bauswein2013,Hotokezaka2013,Lehner2016,Dietrich2017}. For $q> 0.8$, we interpolate $f_{\rm red}$ between the available simulations using a smooth polynomial fit (shown in appendix Figure \ref{fig:ye}). The predicted ejecta masses in each component, for a range of NS binaries, are shown in Figure \ref{fig:mass}. Following \citet{Radice2018}, we assume a fixed grey opacity for each component: $\kappa_{\rm blue}=0.5$\,cm$^2$\,g$^{-1}$, and $\kappa_{\rm red}=10$\,cm$^2$\,g$^{-1}$ \citep{Villar2017,Metzger2019,Tanaka2020}.

\subsection{Post-merger ejecta}

For most systems, additional mass will be ejected after the merger, driven by neutrino-heated winds from a viscous accretion disk around the remnant or from the surface of a massive remnant NS \citep{Metzger2014}. Simulations show that in many cases the mass of post-merger ejecta will exceed that of the dynamical ejecta \citep{Ciolfi2017,Siegel2017,Siegel2018,Radice2018,Fernandez2019}.

The properties of the wind ejecta depend on the lifetime of the merger remnant, which may be a stable NS, a supramassive (hypermassive) NS temporarily supported by (differential) rotation, or may promptly collapse to a BH. To first order, longer-lived remnants produce a larger disk mass with a higher $Y_e$ \citep{Metzger2014,Lippuner2017}.

We use the analytic expression for the disk mass from \citet{Coughlin2019}, who determine the values of their coefficients $a,b,c,d$ by fitting to simulations from \citet{Radice2018}:
\begin{equation}
\label{eq:disk}
    \log_{10}(M_{\rm disk}) = {\rm max}\left(-3, a\left(1+b \tanh{\left(\frac{c-M_{\rm tot}/M_{\rm thr}}{d}\right)}\right)\right).
\end{equation}
Here $M_{\rm tot} = M_1+M_2$ is the total binary mass, and $M_{\rm thr}$ is the threshold mass for prompt collapse to a BH, parameterised as 
\begin{equation}
\label{eq:thr}
    M_{\rm thr} = \left(2.38 - 3.606 \frac{M_{\rm TOV}}{R_{1.4}}\right)M_{\rm TOV},
\end{equation}
i.e.~proportional to the maximum stable mass of a non-rotating NS and the compactness \citep{Bauswein2013b}. An alternative fit by \citet{Radice2018} was parameterised in terms of the tidal deformability rather than $M_{\rm TOV}$. Following \citet{Coughlin2019}, we prefer to retain the explicit dependence on $M_{\rm TOV}$ so that we can try to constrain it observationally. Recent fits by \citet{Nedora2020} predict disk masses broadly consistent with the results of \citet{Coughlin2019}.
The parameterisation above is based on simulations of systems with close to equal mass ratio, $q \approx 1$ \citep{Bauswein2013,Radice2018}. We note however that \cite{Kiuchi2019} have shown that the mass ratio can play an important role and enhance the disk mass for small $q$.

A fraction $\epsilon_{\rm disk}\sim 0.1-0.5$ of the disk mass is expected to be ejected by viscously-driven winds, with $\epsilon_{\rm disk}\approx 0.2$ likely typical \citep{Siegel2017,Radice2018}. This fraction can be fixed in our model for population synthesis, or left free to infer its value from fits to observed KNe
(although in principle, $\epsilon_{\rm disk}$ may change as a function of disk mass; \citealt{De2020}).
Additional post-merger ejecta may arise through neutrino- or magnetically-driven winds \citep{Radice2018,Metzger2018,Ciolfi2020}, which we will discuss in section \ref{sec:ext}.

To determine the velocity and composition of the disk ejecta, we first estimate the remnant properties as a function of $M_{\rm tot}$, $M_{\rm TOV}$ and $M_{\rm thr}$ using table 3 of \citet{Metzger2019}. For $M_{\rm tot}< M_{\rm TOV}$, the remnant is indefinitely stable and we assume an ejecta velocity of $\approx 0.1c$ based on the simulations of \citet{Metzger2014}. For $M_{\rm tot} > M_{\rm thr}$ (prompt collapse), the same simulations predict a velocity of $\approx 0.03c$. We linearly interpolate between these velocities for intermediate $M_{\rm tot}$. Figure \ref{fig:mass} shows the predicted disk ejecta masses and velocities alongside the dynamical ejecta from section \ref{sec:dyn}.

To estimate the opacity we use the simulations from \citet{Lippuner2017}, who present $Y_e$ distributions in their Figure 2 for disks with various remnant lifetimes, ranging from indefinitely stable to prompt collapse. We integrate these distributions to determine the mass-weighted $Y_e$. We then convert this mean $Y_e$ into an effective grey opacity, $\kappa_{\rm purple}$, using fits to the $Y_e - \kappa$ relations from \citet{Tanaka2020} (shown in appendix Figure \ref{fig:kappa}). The inferred range of $0.25<\langle Y_e \rangle < 0.38$ corresponds to an opacity range of $5.5\gtrsim \kappa_{\rm purple}/({\rm cm}^2{\rm g}^{-1})\gtrsim1.5$ for the disk ejecta. The `purple' subscript refers to the fact that this lies intermediate between the red and blue dynamical ejecta.

\subsection{Luminosity from r-process decay}

A semi-analytic model that captures the essential light curve physics of kilonovae was developed by \citet{Metzger2019}. This model approximates the heating rate at time $t$, averaged over all r-process radioactive decay chains, as $\dot{Q}(t)\propto \epsilon_{\rm th} M_{\rm r} t^{-1.3} $ \citep{Korobkin2012}, where $M_{\rm r}$ is the r-process mass and $\epsilon_{\rm th}$ is a time-dependent thermalisation efficiency \citep{Barnes2016}. A module to calculate $\dot{Q}(t)$ in this way was implemented in \textsc{mosfit} by \citet{Cowperthwaite2017} and \citet{Villar2017}.

To determine the output luminosity, $L(t)$, the  ejecta are assumed to have a constant grey opacity and expand homologously with scale velocity $v_{\rm ej}$, allowing for an analytic radiative diffusion solution to the first law of thermodynamics \citep{Arnett1982}. The temperature of the photosphere is initially calculated using the Stefan-Boltzmann law, with radius $R=v_{\rm ej} t$, until such time as the cooling ejecta start to recombine. At this point, the photosphere recedes into the ejecta while maintaining the recombination temperature, $T_{\rm rec}$. The \textsc{mosfit} modules to solve for the radiative diffusion and the photospheric temperature and radius are described in \citet{Nicholl2017b}. 

\citet{Cowperthwaite2017} and \citet{Villar2017} generalised this model to include the sums of multiple ejecta components: the luminosity scale of each component is proportional to its mass, with a diffusion timescale determined also by its velocity and opacity. They allowed $T_{\rm rec}$ to vary independently for each of 2-3 ejecta components, finding best fit values in the range $1000\,{\rm K}\lesssim T_{\rm rec} \lesssim 4000\,{\rm K}$. To reduce the number of free parameters (and restrict to binary-based quantities of interest) we assume a fixed $T_{\rm rec}= 2500$\,K, based on the theoretically-predicted recombination temperature of ionised r-process ejecta \citep{Barnes2013}. This falls in the middle of the plausible range found by \citet{Villar2017}. We tested the effect of leaving this parameter free, and found a best-fit value $T_{\rm rec, free}\approx 2600\pm200$\,K, consistent with theory. For completeness, we also tested the effect of using three independent temperatures in our base model; in this case they converged to similar values to \citet{Villar2017}, and other parameters remained consistent with the fixed temperature results to better than $1\sigma$.

\begin{figure*}
    \centering
    \includegraphics[width=15cm]{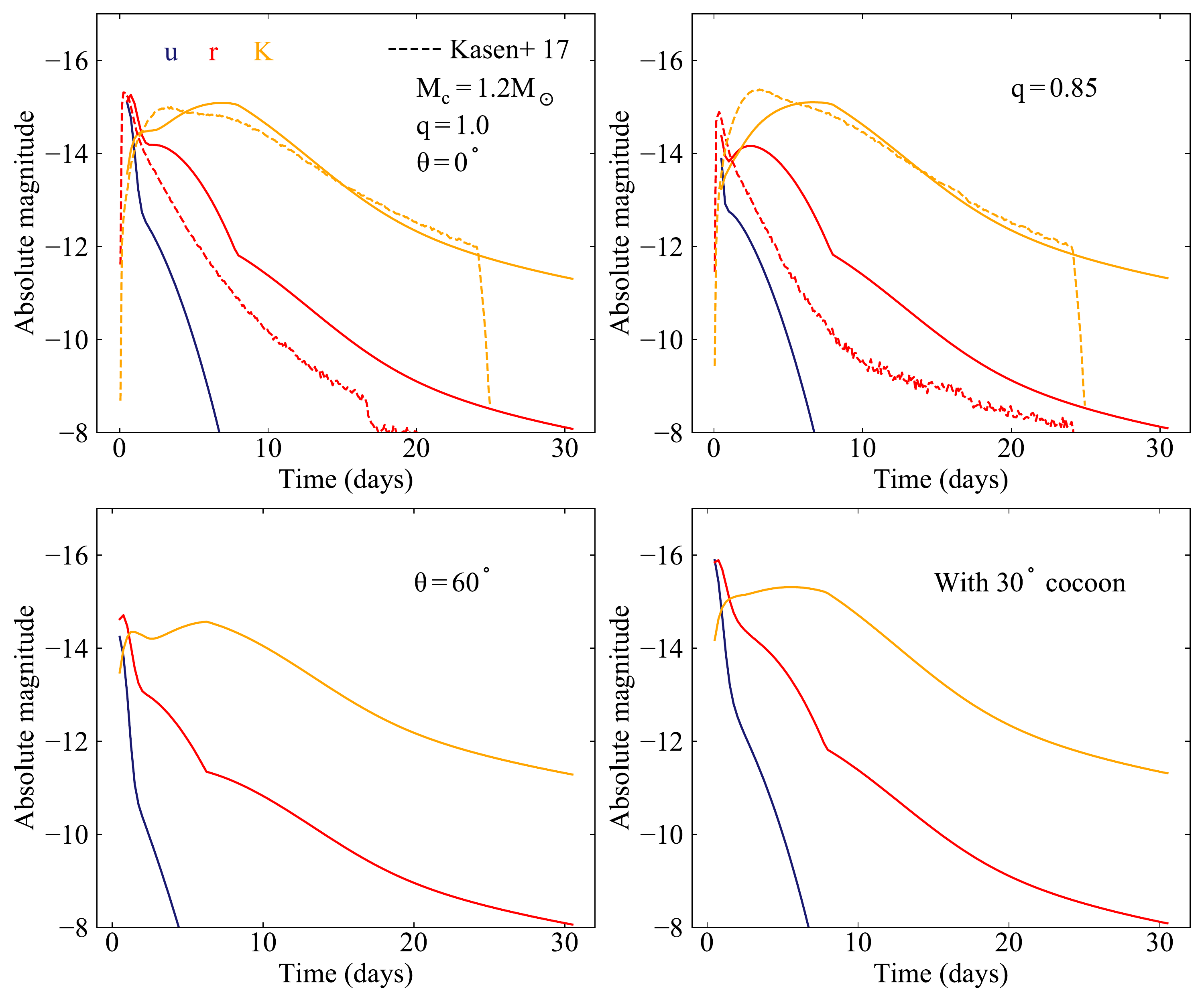}
    \caption{Example light curves from our model in the near-UV ($u$), optical ($r$) and near-infrared ($K$) bands, for a merger with chirp mass $\mathcal{M}=1.2$\,\M. Top left: Fiducial, equal mass merger viewed along pole. Top right: Merger with same chirp mass but asymmetric mass ratio, $q=0.85$. Lower left: Fiducial model viewed at 60$^\circ$ off-axis. Lower right: Including shock-heated cocoon with opening angle 30$^\circ$. The top two panels show light curves calculated with the models of \citet{Kasen2017} for the closest available ejecta masses (assuming lanthanide fractions of $10^{-9}$, $10^{-2}$ and $10^{-1}$ for the blue, purple and red components, respectively). These models have not been tuned to match our simulations and are intended only to guide the eye. The viewing angle effect is very roughly approximated by adding $-0.5$\,mag to the $r$ band and $+0.5$\,mag to the $K$ band.}
    \label{fig:examples}
\end{figure*}

\subsection{Geometry} 

Most analytic light curve models assume spherical symmetry. However, in kilonovae the separate ejecta components occupy distinct spatial regions, due to their different ejection mechanisms (as illustrated in Figure \ref{fig:schem}). The low-medium $Y_e$ (i.e.~blue and purple) ejecta are expected to escape mostly orthogonal to the orbital plane, where shock heating and weak interactions lower the free neutron flux \citep[e.g.][]{Wanajo2014,Goriely2015,Foucart2016}. In contrast, the neutron-rich red ejecta escape parallel to the orbital plane since they form tidally. Simulations show that this geometry can be well approximated as a sphere with conical sections of half-opening angle $\theta_{\rm open}\sim 45^\circ$ around the poles. The blue/purple ejecta reside within the conical sections and the red ejecta outside.

This results in a viewing angle dependence in the observed luminosity, colour and timescale of the kilonova \citep[e.g.]{Kasen2015,Kawaguchi2018,Wollaeger2018,Bulla2019,Korobkin2020}, which can be understood intuitively: from viewing angles close to the equator, emission from the low-lanthanide polar ejecta is obscured by the curtain of lanthanide-rich ejecta at lower latitudes, meaning that only a slow-evolving red kilonova is observed. Radiative transfer simulations show that this can reduce the peak luminosity by $\approx 1-2$ magnitudes \citep{Bulla2019}. Neglecting this effect is problematic both for population synthesis and for fitting observed kilonovae, since an offset in the luminosity introduces a systematic error in measuring the ejected mass.
Another subtle viewing angle effect is associated with the polar cavity carved out by a GRB jet (whether failed or successful) that can accompany the merger, as recently studied by \cite{Nativi2021,Klion2021}. This can expose interior regions of the ejecta which are hotter or lower opacity, resulting in brighter or bluer emission.

\citet{Villar2017} approximated the viewing angle effect in \textsc{mosfit} by multiplying the luminosity of the equatorial ejecta by $\cos{\theta}$ and the polar ejecta by $1-\cos{\theta}$, with the viewing angle $\theta$ a free parameter. Here we implement a new module that treats this more exactly, using the formalism of \citet{Darbha2020}. This takes a sphere with conical polar caps of half-opening angle $\theta_{\rm open}$, and scales the luminosity of each component using the projected area of the caps (blue/purple ejecta) or remaining sphere (red ejecta) as a function of $\theta$ and $\theta_{\rm open}$. We have verified that the change in luminosity with $\theta$ is consistent with the radiative transfer results of \citet{Bulla2019} for the same geometry. 
We note that this treatment does not account for the jet-interaction viewing angle effect discussed by \cite{Klion2021,Nativi2021}, as this would depend on uncertain properties of the jet.
For our fiducial model, we assume $\theta_{\rm open}=45^\circ$ and vary the observer viewing angle $\theta$.

\subsection{Extensions to the base model}
\label{sec:ext}

We include in our model two optional final ingredients that can contribute additional luminosity. Although they are tied less explicitly in our formalism to the initial binary parameters, neglecting these contributions may lead to an underestimate in the model luminosity or to a bias in the inference of other parameters from kilonova data. These components can easily be turned off at run-time, allowing the user to recover the fully predictive baseline model described in the previous sections.

The first is an enhancement in the blue ejecta due to magnetically-driven winds \citep{Metzger2018,Ciolfi2020}. This mechanism is possible only if the remnant avoids prompt collapse, hence we apply this effect only when the merger mass is below $M_{\rm thr}$ (equation \ref{eq:thr}). We adopt a simple parameterisation, inspired by \citet{Coughlin2019}: $M_{\rm blue,tot} = M_{\rm blue,dyn}/\alpha$, where $0<\alpha<1$. Therefore fixing $\alpha=1$ turns off this source of ejecta.

The second effect is shock-heating of the ejecta by a GRB jet. GW170817 was accompanied by a short GRB (as expected in NS mergers; \citealt{Eichler1989,Narayan1992,Berger2014}), and some studies interpreted the early blue luminosity of the kilonova as arising from cooling of this shocked `cocoon' \citep{Kasliwal2017,Arcavi2018}. We include this effect using the formulae from \citet{Piro2018}, specifically their equations 38 and 39. The breakout radius is approximated as the time delay between the GW and GRB signals \citep[1.7\,s for GW170817;][]{Abbott2017b} multiplied by the speed of light \citep{Piro2018}. Following \citet{Nakar2017}, we assume that the shock deposits constant energy per decade of velocity in the ejecta. The isotropic-equivalent luminosity is proportional to the mass of shocked ejecta, which is a fraction $\theta_{\rm c}^2/2$ of the total ejecta. The shock can therefore be suppressed by setting the cocoon half-opening angle $\theta_{\rm c}$ equal to zero (or $\cos{\theta_{\rm c}}=1$). We assume that the shocked mass is a fraction of the total dynamical ejecta, as it is unclear whether post-merger ejecta will precede the launching of the GRB jet. The output light curves are only weakly sensitive to this assumption, partially due to degeneracy between $M_{\rm ej}$ and $\theta_{\rm c}$. Because the cocoon shock occurs primarily in the polar material, we assume this matter has the same opacity as the blue ejecta component. The luminosity of the shock is inversely proportional to this opacity, so we note that a lanthanide-rich polar component could significantly reduce the importance of the shock emission.

We show examples of light curves calculated for different mass ratios, inclinations and cocoon opening angles in Figure \ref{fig:examples}.

\begin{table*}
     \caption{Parameters in the \textsc{mosfit} \textsc{bns} model and application to GW170817. Some parameters are relevant only when fitting to data. Priors in square brackets are flat over the specified ranges, otherwise Gaussian. Posterior values for tidal deformability are given in terms of $\tilde{\Lambda}$ to facilitate comparison to GW analyses. The `Surface ejecta' model includes additional blue ejecta from a long-lived remnant; the `Shocked cocoon' model includes emission from a GRB-heated cocoon; the `Agnostic' model includes both. The stated best-fit values and uncertainties correspond to the medians and 16th/84th percentiles of the marginalised posteriors.}
   \begin{center}
    \begin{tabular}{lllllll}
        & \multicolumn{2}{c}{Prior distribution} & \multicolumn{4}{c}{Posterior distribution for fit to GW170817} \\ 
        Parameter  & Default prior & GW170817 prior & Base model & Surface ejecta  & Shocked cocoon  & Agnostic \\
       \hline
        $\mathcal{M}$$^a$ (\M) & $[0.7,2.0]$ & $1.188\pm0.002$ & $1.188\pm0.002$ & $1.188\pm0.002$ & $1.188\pm0.002$ & $1.188\pm0.002$ \\
        $q$$^b$ & $[0.4,1.0]$ & $[0.7,1.0]$ & $0.99\pm0.01$ & $0.88^{+0.04}_{-0.02}$ & $0.95^{+0.04}_{-0.07}$ & $0.92\pm0.05$ \\
        $\cos{\theta}$$^c$ & $[0.0,1.0]$ & $ [0.82,0.97]$ & $0.89\pm0.04$ & $0.88^{+0.04}_{-0.02}$ & $0.85^{+0.03}_{-0.02}$ & $0.84^{+0.03}_{-0.02}$\\
        $M_{\rm TOV}$$^d$ (\M) & $[2.0,2.5]$ & $[2.0,2.5]$ & $2.32\pm0.04$ & $2.26\pm0.03$ & $2.14^{+0.06}_{-0.07}$ & $2.17^{+0.05}_{-0.06}$ \\
        $\Lambda_{\rm s}$$^e$ & $[10,2000]$ & $[10,2000]^{*}$ & $\tilde{\Lambda}=122^{+15}_{-13}$ & $\tilde{\Lambda}=116^{+24}_{-12}$ & $\tilde{\Lambda}=163^{+57}_{-44}$ & $\tilde{\Lambda}=231^{+92}_{-74}$ \\
        $\epsilon_{\rm disk}$$^f$ & $[0.05,0.5]$ & $[0.05,0.5]$ & $0.14\pm0.01$ & $0.13\pm0.01$ & $0.12\pm0.01$ & $0.12\pm0.01$ \\
        $\alpha$$^g$ & $[0.1,1.0]$ & $[0.1,1.0]$ & --- & $0.37^{+0.17}_{-0.15}$ & --- & $0.63^{+0.18}_{-0.19}$ \\
        $\cos{\theta_{\rm c}}$$^h$ & $[0.707,1.0]$ & $[0.707,1.0]$ & --- & --- & $0.89^{+0.04}_{-0.06}$ & $0.91^{+0.04}_{-0.07}$ \\
        $t_0$$^i$ (days) & $[-1.0,0.0]$ & ($-0.462$) & --- & --- & --- & --- \\
        $\log{N_{\rm H}}^j$ & $[19.0,23.0]$ & $[19.0,20.3]$ & $19.35^{+0.31}_{-0.22}$ & $19.55^{+0.41}_{-0.34}$ & $19.65^{+0.39}_{-0.40}$ & $19.67^{+0.39}_{-0.41}$ \\
        $\log{\sigma}^k$ & --- & $[-3.0,2.0]$ & $-0.34\pm0.02$ & $-0.39\pm0.02$ & $-0.44\pm0.02$ &  $-0.44\pm0.02$ \\
        \hline
         & \multicolumn{2}{l}{Bayesian model evidence ($\ln{Z}$):} &  37.1 &  56.9 &  80.3 &  81.1 \\
        \hline
    \end{tabular} \\
    \end{center}
    $^a$~Chirp mass; $^b$~Mass ratio; $^c$~Observer viewing angle; $^d$~Maximum stable NS mass $^e$~Symmetric tidal deformability (specified instead as $R_{1.4}$ in generative mode); $^f$~Fraction of disk ejected; $^g$~Enhancement ($1/\alpha$) of blue ejecta by surface winds; $^h$~Opening angle of shocked cocoon; 
    $^i$~Time of explosion (rest-frame);
    $^j$~Hydrogen column density in host galaxy (proportional to extinction);
    $^k$~White noise in likelihood function; $^{*}$Additional constraint that derived parameter $\tilde{\Lambda}<800$, see section \ref{sec:gw} \\  
    \label{tab:params}
\end{table*}

\subsection{Free parameters and relation to GW observables}
\label{sec:gw}

After implementing the modules described above, our model depends on the masses of the constituent neutron stars, their compactness, the maximum stable mass $M_{\rm TOV}$, and the viewing angle $\theta$. The fraction of the disk mass ejected in the purple component can be varied or fixed to a fiducial value. Two further parameters allow one to increase the early blue flux through magnetic winds and/or shock cooling. Finally, we include line-of-sight extinction, and a white noise parameter $\sigma$ in our likelihood function when fitting to data \citep[see][]{Nicholl2017b,Villar2017,Guillochon2018}.

There is a choice to be made about how best to express some of these parameter dependencies. We opt to use the parameterisation that most closely matches the quantities constrained by GW observations. This allows the user to quickly predict the electromagnetic light curve for a given GW signal, or to use GW information to inform their priors when fitting to an observed multi-messenger source. We therefore parameterise the masses of the system in terms of the `chirp' mass $\mathcal{M}=(M_1 M_2)^{3/5}(M_1+M_2)^{-1/5}$, the quantity most accurately measured from the GW signal, and $q=M_2/M_1 \le 1$, to which this signal is also somewhat sensitive.

The dependence on the NS compactness $C$ is more complicated. For population synthesis, it is simplest to specify a radius $R$, equivalent to implicitly choosing an equation of state (since all NSs of comparable mass should have approximately the same radius). However, when fitting to observed data (i.e. when we want to measure, rather than impose, a NS radius), it is more useful to express $C$ in terms of the NS tidal deformability, which measures the responsiveness of the NS to an external tidal field: $\Lambda_i = (2/3)k_2 C_i^{-5}$, where $k_2(M)\approx 0.05-0.15$ is the quadrupolar tidal Love number \citep{Hinderer2008,Postnikov2010,Damour2012}. 

The reason is that the observable GW phase is to leading order determined by a specific combination of $\Lambda_1$ and $\Lambda_2$ \citep{Flanagan2008,Wade2014,Favata2014},
\begin{equation}
\label{eq:lam}
    \tilde{\Lambda}=\frac{16}{13}\frac{(m_1+12M_2)M_1^4\Lambda_1+(M_2+12M_1)M_2^4\Lambda_2}{(M_1+M_2)^5},
\end{equation}
called the binary or effective tidal deformability, which can therefore be constrained with GW data. 
The relation between $\Lambda$ and $C$ allows us to include GW information in our priors to better measure $C$ and hence $R$. A similar approach was adopted by \citet{Coughlin2019}, though they used a different relation proposed by \citet{De2018} to relate $\tilde{\Lambda}$ to $R$ without calculating $C$ explicitly.

For consistency with GW analyses \citep{Abbott2018} and to reduce systematic error, we instead sample $\Lambda$ in terms of the so-called `symmetric' tidal deformability:
\begin{equation}
\label{eq:sym}
    \Lambda_{\rm s} \equiv (\Lambda_1+\Lambda_2)/2.
\end{equation}
Using `quasi-universal relations' or QURs (largely independent of the EoS, and hence of complicating factors such as $k_2$), the antisymmetric tidal deformability $\Lambda_{\rm a}\equiv (\Lambda_2-\Lambda_1)/2$ can be derived from $\Lambda_{\rm s}$ to $\approx 5\%$ precision \citep{Yagi2016sym}, allowing one to reconstruct $\Lambda_1$, $\Lambda_2$ and $\tilde{\Lambda}$ directly from $\Lambda_{\rm s}$. One may then impose the constraint that $\tilde{\Lambda}$ be consistent with the GW results, using the \textsc{constraints} class in \textsc{mosfit} \citep[see][]{Guillochon2018}. The compactness of each NS is then derived from $\Lambda_i$ using another QUR:
\begin{equation}
\label{eq:yagi}
    C_i = 0.360 - 0.0355\ln{(\Lambda_i)} + 0.000705\ln{(\Lambda_i)}^2
\end{equation}
This relation is accurate to $\approx 6.5\%$ across a variety of equations of state \citep{Yagi2017com}. These $C_i$ are then used in equation \ref{eq:Mdyn} to estimate the dynamical ejecta mass. Following a fit to observed data, the NS radii can be derived from the posterior distribution of $\Lambda_{\rm s}$ following the same procedure and using the constituent NS masses in equation $\ref{eq:C}$.

Our final parameter set therefore consists of 5-11 parameters, which differ slightly between the generative and fitting versions of the model. We list all of these parameters and suggest priors in Table \ref{tab:params}. For comparison, the parameterisation of the original \textsc{mosfit} kilonova model by \citet{Villar2017} also had 11 free parameters.

\begin{figure*}
    \centering
    \includegraphics[width=\columnwidth]{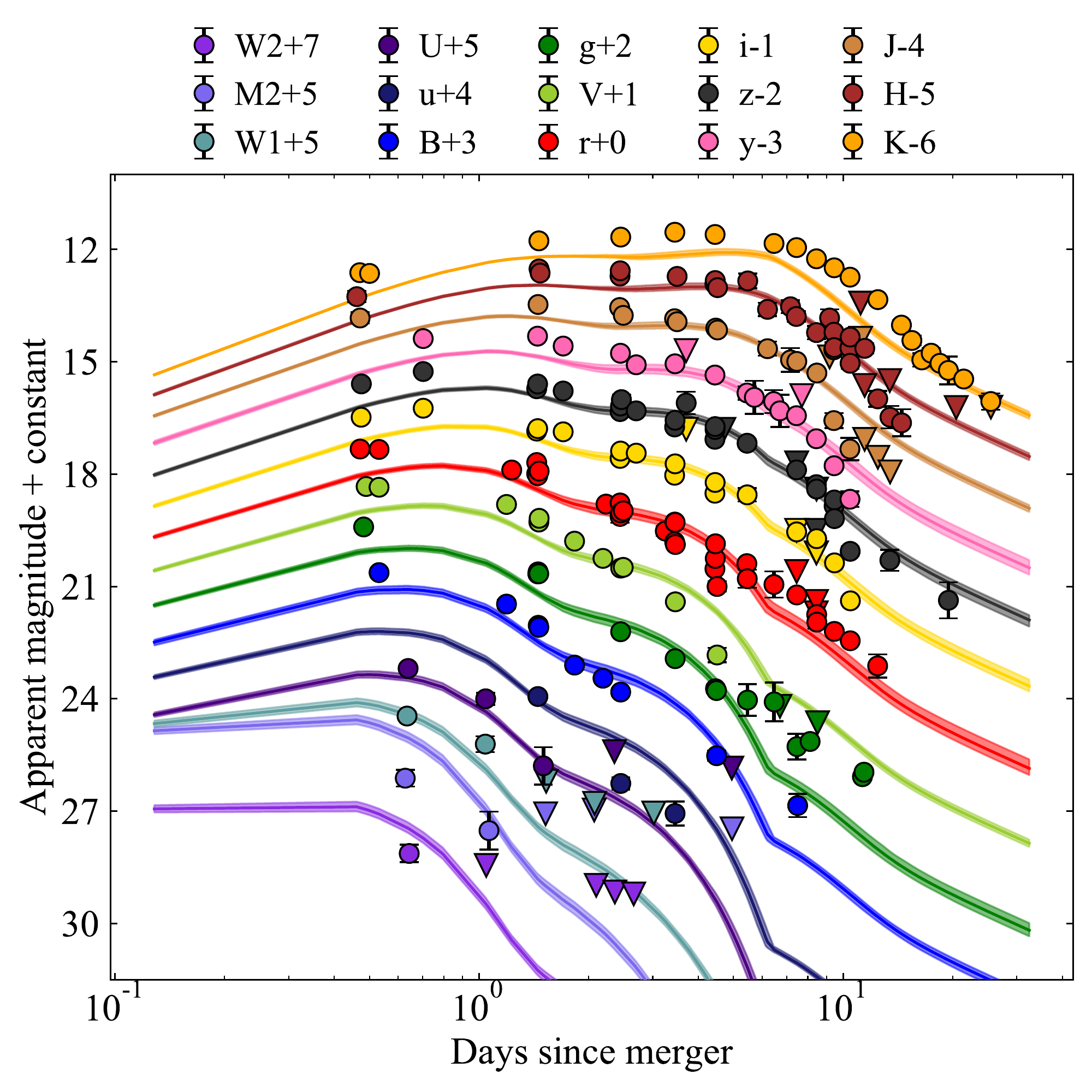}
    \includegraphics[width=\columnwidth]{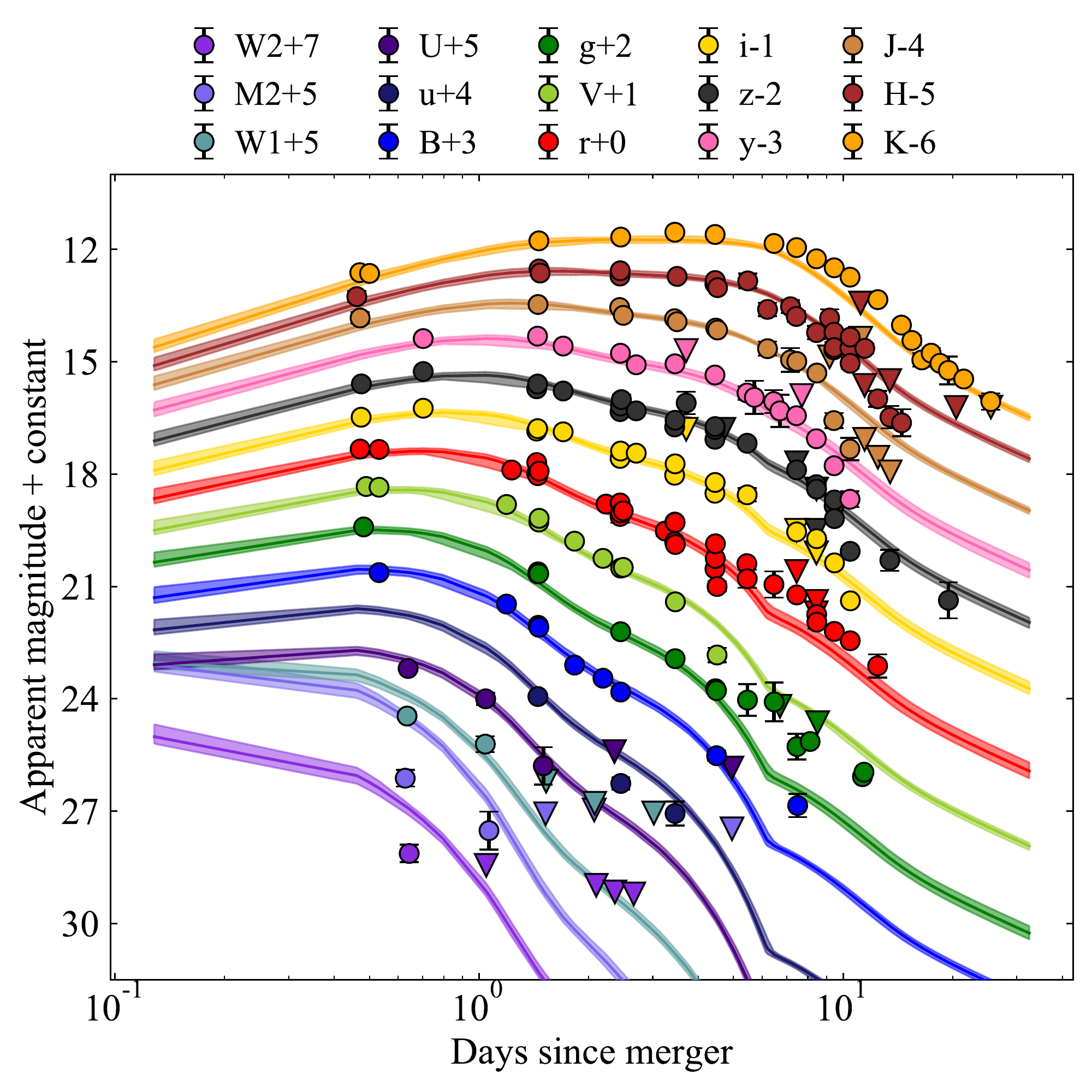}
    \caption{Left: Best-fit kilonova model for GW170817 without surface ejecta enhancement or GRB shock cooling. Right: Best-fit model with these effects included. Both models provide a good match to the data beyond $t\gtrsim 1.5$ days post-merger, but the fit without shock cooling is too faint by $\approx 0.7$ magnitudes (almost a factor 2) during the first day. The model on the right is preferred to the base model with a Bayes factor $B>10^{19}$. Light curve fits for the surface ejecta and shock-only models are shown in the appendix.}
    \label{fig:lc}
\end{figure*}

\begin{figure*}
    \centering
    \includegraphics[width=\textwidth]{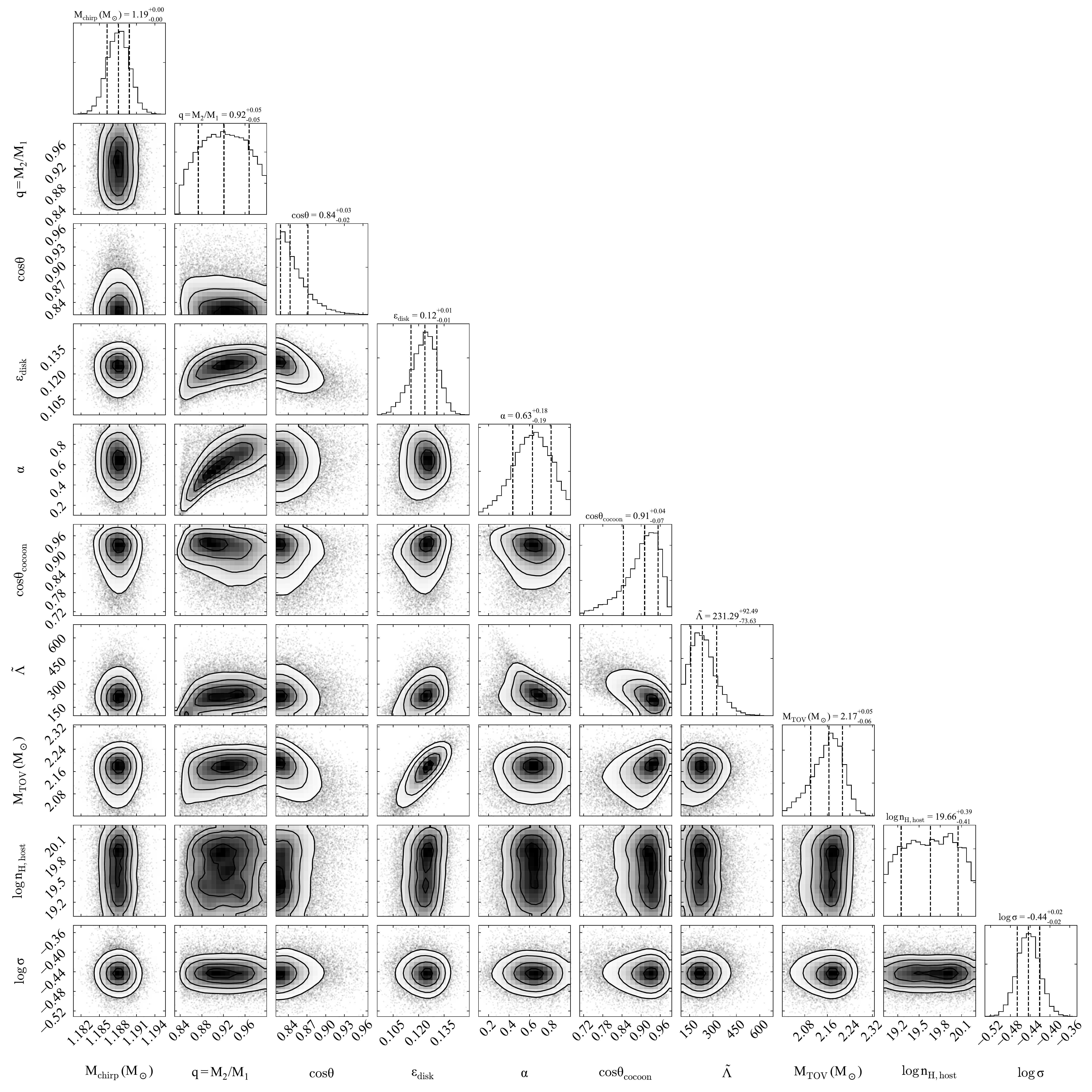}
    \caption{Posterior distributions of free parameters for the preferred agnostic model (including shock cooling and enhanced surface wind ejecta) fit to GW170817, using multi-messenger priors (Table \ref{tab:params}).}
    \label{fig:cornershock}
\end{figure*}

\section{Application to GW170817}
\label{sec:170817}

As with any model of NS mergers, GW170817 provides an ideal testing ground for the formalism described above. In this section, we fit the observed optical data to demonstrate (i) how using a binary (rather than ejecta) based parameter set provides additional insight into the physical origin of each luminosity component; (ii) the utility of this model in probing the NS EoS with kilonova data; and (iii) the extent to which the use (or not) of GW information in the model priors affects our posteriors.

We use data from the Open Kilonova Catalog \citep{Guillochon2017}, compiled by \citet{Villar2017}. Given the high level of overlap between the observations from different groups, we fit the following subset, chosen to cover the full range of bands with a comparable (approximately nightly) density of sampling in each: $ugrizyH$ from \citet{Cowperthwaite2017}, $BVgrJH$ from \citet{Drout2017}, $rizy$ from \citet{Smartt2017}, $rzK$ from \citet{Tanvir2017}, $JHK$ from \citet{Kasliwal2017}, $BV$ from \citet{Troja2017}, and the $U$ and UV bands from \citet{Evans2017}. 

Our priors are given in Table \ref{tab:params}. The time of merger is fixed by the GW detection. The priors on $\mathcal{M}$ and $q$ are from \citet{Abbott2017a}, while the prior on $\Lambda_{\rm s}$ is chosen to match the analysis employed by \citet{Abbott2018}. The range of viewing angles, $0.82<\cos{\theta}<0.97$, is from \citet{Nakar2020}, who compiled results from GRB afterglow modelling \citep{Alexander2017,Alexander2018,Haggard2017,Margutti2017,Margutti2018,Troja2017,Troja2018,Troja2019,DAvanzo2018,Dobie2018,Gill2018,Granot2018,Lazzati2018,Lyman2018,Mooley2018,Fong2019,Hajela2019,Lamb2019,Wu2019,Ryan2020} and from very long baseline interferometry \citep[VLBI;][]{Ghirlanda2019,Hotokezaka2019}. We assume a distance of 40.7\,Mpc \citep{Cantiello2018} to the host galaxy, NGC\,4993. 

We run our fits on the University of Birmingham \textsc{bluebear} cluster, and use \textsc{dynesty} \citep{Speagle2020} to integrate the model evidence and sample the posteriors of the model parameter space. We use the likelihood function
\begin{equation}
    \ln\mathcal{L}=-\frac{1}{2}\sum_{i=1}^{n}\left[\frac{(O_i-M_i)^2}{\sigma_i^2+\sigma^2}-\ln(2\pi\sigma_i^2)\right]-\frac{n}{2}\ln(2\pi\sigma^2),
\end{equation}
where $O_i$ and $M_i$ are the set of observed and model magnitudes, $\sigma_i$ are the errors on the data, and $\sigma$ is a white noise free parameter to account for additional uncertainty in the data or model; for a good fit within the observed errors, one therefore finds $\sigma \lesssim\ \langle\sigma_i\rangle$.
\citet{Speagle2020} gives an extensive account of how the evidence is evaluated directly in \textsc{dynesty}; we refer the reader there for details (and a good overview of Bayes' Theorem). In short, random draws from the model priors are used to integrate numerically the hypervolume between shells of constant likelihood. The products of likelihood and volume for these points yield a set of weights that can be summed to determine the total evidence. Points are evolved by replacing those with lowest likelihood; the weights of the final set are proportional to the model posterior \citep{Speagle2020}.

\subsection{Model comparison and importance of shock cooling}

We begin by fitting GW170817 with four variants of the model to determine the relative importance of different effects. These are (i) the baseline model ($\alpha=\cos{\theta_{\rm c}}=1$), (ii) a model with blue ejecta enhancement from e.g.~magnetic surface winds ($\alpha\leq 1$), (iii) a model with shock-cooling of a GRB cocoon ($\cos{\theta_{\rm c}}\leq 1$), and (iv) an agnostic model allowing for both effects ($\alpha,\cos{\theta_{\rm c}}\leq 1$). The model evidence and marginalised posteriors are given in Table \ref{tab:params}. We show resulting light curves compared to the GW170817 data in Figure \ref{fig:lc}.

We use the Bayes factor, $B\equiv Z_1/Z_2$ where $Z_i$ is the Bayesian evidence for model $i$, to compare these models. Generally, $B>10$ is taken to indicate a strong preference, based on the available data, for one model over another, and $B>100$ to indicate a decisive preference. Compared to the base model, the model with enhanced surface wind ejecta is preferred with $B>10^8$. 

However, \emph{our fits overwhelmingly prefer models that include shock cooling} compared to either of the other models, with Bayes factors $B>10^{10}$ compared to the surface ejecta model and $B>10^{19}$ compared to the base model. The agnostic model (with both surface winds and shocks) is weakly favoured over the shock-only model, with $B=2.2$. The decisive preference for shock cooling can be understood from Figure \ref{fig:lc}. Models without shocks, generating their early luminosity only from radioactive decay in the blue ejecta, are too faint by $\approx 0.7$ magnitudes (almost a factor 2 in luminosity) at $t\sim0.5$ days. Increasing the blue ejecta mass through the parameter $\alpha$ improves the fit at $t\sim 1.5$ days, but is still too faint during the first night (see appendix). However, a cocoon with an opening angle $\theta_{\rm c}\sim 20-30^\circ$ provides the excess luminosity at $t=0.5$ days required to match the data.
This supports previous work that has argued for the importance of shock cooling in the early light-curve \citep{Kasliwal2017,Gottlieb2018,Piro2018}. 

As an important caveat, we note that there are other models that may be degenerate with cocoon emission. Free neutron decays, if present, could heat the outermost ejecta during the first few hours \citep{Kulkarni2005,Metzger2015}, and produce a signal degenerate with shock heating \citep{Metzger2018}. \cite{Nativi2021,Klion2021} recently showed that the bright early light-curve can alternatively be explained by accounting for an asymmetric ejecta distribution induced by the GRB jet. Finally, any differences in the nuclear heating rates from those assumed here would also affect the peak luminosity, as recently shown by \citet{Zhu2021}. These effects are not modeled within our Bayesian analysis, and therefore remain a potential alternative to the shock-cooling interpretation.

\subsection{Fit parameters and the origin of each luminosity component}

Taking the agnostic version as our preferred model, we show the two-dimensional posteriors for this fit in Figure \ref{fig:cornershock}. As with all model variants considered, the chirp mass posterior is essentially the prior, since this parameter is constrained to such high precision by the GW data. The electromagnetic data tighten the mass ratio of the system to $q>0.84$ (about 50\% of the GW-inferred prior volume) due to the requirement for blue ejecta. Viewing angles close to $\sim 30^\circ$ are preferred, more in line with afterglow modelling than with VLBI \citep{Hajela2019,Nakar2020}. This is also consistent with the GW-inferred viewing angle at the distance of the GW170817 host \citep{Finstad2018}.

The largest degeneracies in the model parameters are between the mass ratio and the surface ejecta enhancement $1/\alpha$, since equal mass binaries also produce more blue ejecta; between \Mtov\ and $\epsilon_{\rm disk}$, since both affect the mass of purple disk ejecta; and between $\theta_{\rm c}$ and $\tilde{\Lambda}$, with a larger shock heating contribution to the luminosity relaxing the preference for low tidal deformability (compact EoS). We will discuss the EoS-dependent quantities, \Mtov\ and \Rns($\tilde{\Lambda}$), in more detail in section \ref{sec:eos}.

The best-fitting models produce typical ejecta masses $M_{\rm blue}\approx0.005$\,\M, $M_{\rm red}\approx0.001$\,\M, and $M_{\rm purple}\approx0.02$\,\M. As we have established, the blue ejecta can only account for $\sim50\%$ of the flux at $t\sim0.5$ days, and the rest is attributed to cooling of a GRB-shocked cocoon. The large mass of purple ejecta dominates the luminosity at $t\sim 2-10$ days, with the low mass of red ejecta contributing significantly only in the extended tail. These results are in agreement with \citet{Villar2017}; however in our forward model we can uniquely associate the purple component with remnant disk winds, the blue component with shock-driven dynamical ejecta (possibly but not necessarily boosted by magnetic surface winds), and the red component with tidal dynamical ejecta, while identifying the importance of the cocoon contribution to the luminosity at early times.

\begin{figure}
    \centering
    \includegraphics[width=\columnwidth]{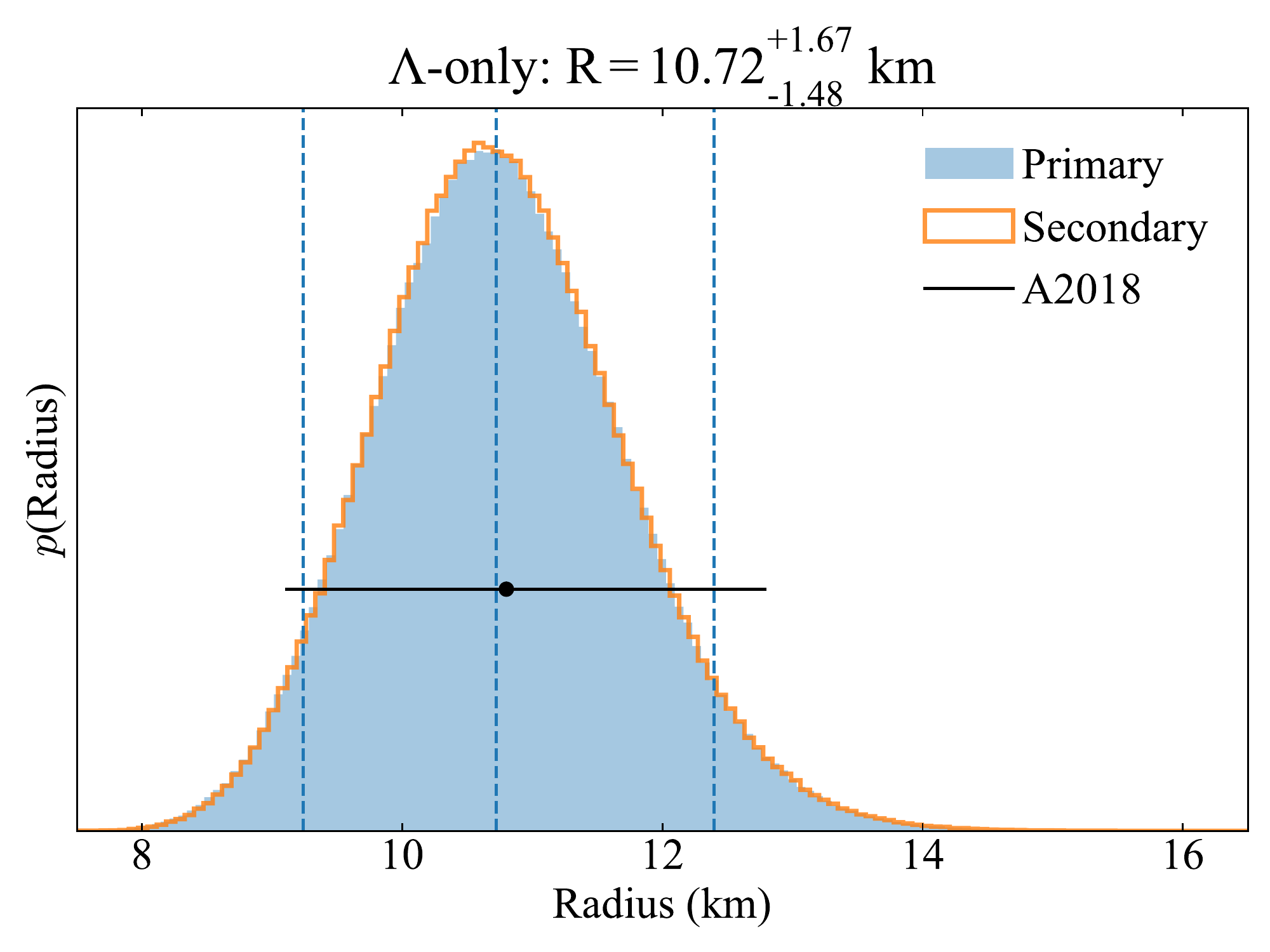}
    \includegraphics[width=\columnwidth]{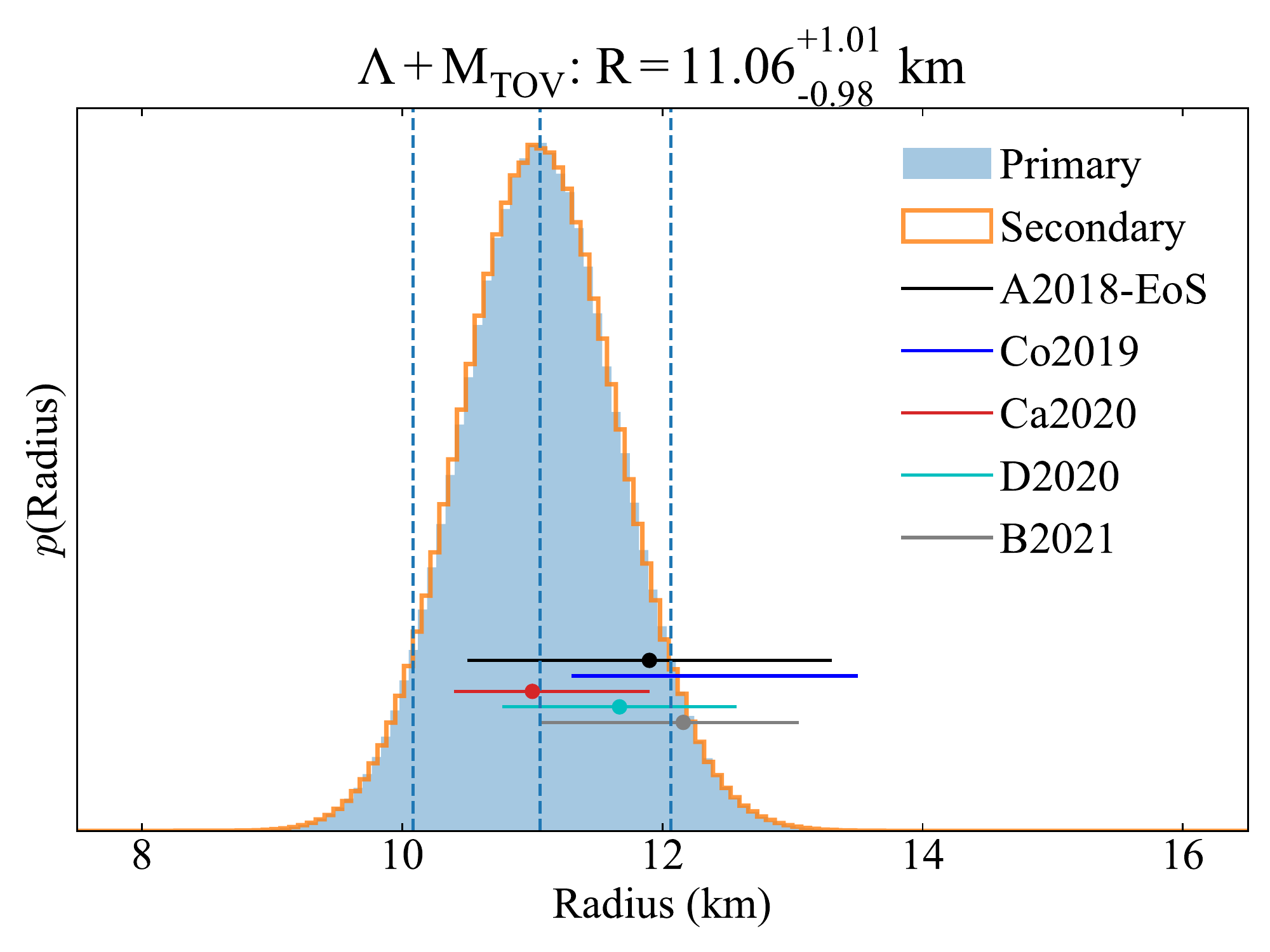}
    \includegraphics[width=\columnwidth]{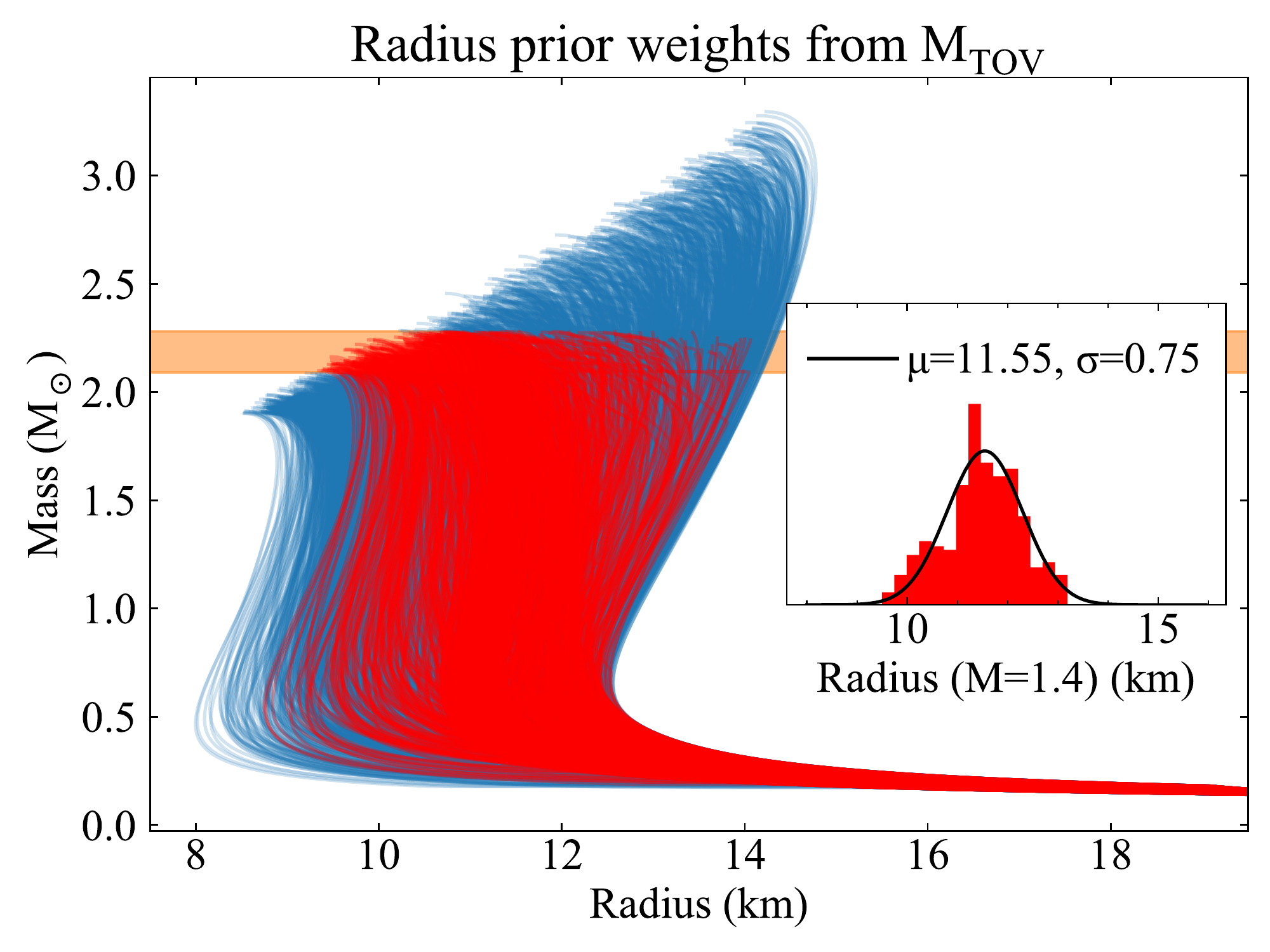}
    \caption{Posterior distributions for the NS radii using the agnostic model, including systematic errors. Vertical lines mark the 5th, 50th and 95th percentiles for the primary (more massive) NS. 
    Plotted for comparison are the measurements by \citet{Abbott2018} (with and without imposing constraints on the EoS), \citet{Coughlin2019}, \citet{Capano2020}, \citet{Dietrich2020} and \citet{Breschi2021}. Top: NS radii from the posterior of $\Lambda_{\rm s}$. Middle: NS radii after re-weighting using equations of state that support an $M_{\rm TOV}$ within our 90\% confidence interval. Bottom: construction of these prior weights using mass-radius curves from \citet{Dietrich2020}. The red curves satisfy our constraint on $M_{\rm TOV}$. Inset: Gaussian fit to allowed radii.}
    \label{fig:radius}
\end{figure}

\subsection{Constraints on the NS equation of state}
\label{sec:eos}

In this section, we examine the posteriors for EoS-dependent quantities in our fit, including estimates of the systematic uncertainty. While \Mtov\ is treated explicitly as a model parameter, the NS radii must be derived from the posterior of the symmetric tidal deformability, $\Lambda_{\rm s}=240^{+94}_{-74}$, and the masses of the two NSs, obtained from $\mathcal{M}$ and $q$.

The major source of systematic uncertainty in \Rns\ and \Mtov\ is the scatter in relations \ref{eq:Mdyn} and \ref{eq:disk}, which are respectively $\sim70\%$ for the dynamical ejecta mass \citep{Dietrich2017} and $\sim50\%$ of our inferred disk mass \citep{Coughlin2019}. We test two methods to simultaneously model these systematic errors. First we run a suite of fits where equations \ref{eq:Mdyn} and \ref{eq:disk} are modified each time by a random factor drawn from a Gaussian distribution with mean equal to 1 and width equal to the calibration uncertainties. Comparing the posteriors of 50 such runs, we find a scatter of $\pm 40$ in $\Lambda_{\rm s}$, and negligible scatter in \Mtov. As a complementary method, we run a single realisation of the fit but with two additional free parameters with Gaussian priors, $\sigma_{\rm dyn}$ and $\sigma_{\rm disk}$ (where the dynamical ejecta mass is $M_{\rm dyn}'=\sigma_{\rm dyn}M_{\rm dyn}$ and the disk ejecta is treated analogously). In the former case, we broaden the posterior of $\Lambda_{\rm s}$ using random draws from a Gaussian of width 40, whereas in the latter case we can use the posterior distribution of $\Lambda_{\rm s}$ directly. We find that the two approaches yield indistinguishable posteriors for $\Lambda_{\rm s}$ (and hence \Rns) and \Mtov.

Since the systematic error is negligible for \Mtov, our estimate of this quantity is equal to the posterior distribution shown in Figure \ref{fig:cornershock}. We measure a 90\% credible interval \Mtov\,$=2.17^{+0.08}_{-0.11}$\,\M. Models without shock cooling (strongly disfavoured by the Bayes factor analysis) prefer a more massive \Mtov\,$\sim2.3$\,\M. For comparison, the analysis of \citet{Margalit2017} found that \Mtov\,$\lesssim2.2$\,\M, which is consistent with our preferred shock cooling models but in potential tension with the shock-free models (although \citealt{Shibata2019b} have extended the \citealt{Margalit2017} analysis and found \Mtov\,$\lesssim 2.3$\,\M).
Other recent works \citep{Rezzolla2018,Lucca2020,Shao2020} have also favoured an EoS with \Mtov\,$\lesssim2.2$\M. It is worth noting that the \Mtov\ constraints obtained here are complementary to these other works, which were based on arguments that are independent from the kilonova modeling adopted here. It is therefore interesting (and non-trivial) that the different approaches yield consistent results.

We next derive \Rns\ from the posteriors of $\Lambda_{\rm s}$, $\mathcal{M}$ and $q$, using equations \ref{eq:C}, \ref{eq:sym} and \ref{eq:yagi}. During this conversion we include the systematic uncertainty introduced by the QURs \citep{Yagi2016sym,Yagi2017com} by adding random scatter to the derived values for $C_i$ (6.5\%) and $\Lambda_{\rm a}$ (5\%), drawn from Gaussian distributions. The masses of the primary and secondary, which enter equation \ref{eq:C}, are $M_1=1.41\pm0.04$\,\M\ and $M_2=1.31\pm0.03$\,\M; therefore their measured radii (especially for the primary) can be taken as very close to \Rns.

The top panel of Figure \ref{fig:radius} shows the derived probability density functions for the NS radii evaluated in this way. We find the primary has a radius $R_1=10.72^{+1.67}_{-1.48}$\,km (90\% credible interval). The models without shock cooling prefer a smaller radius, $\sim 10$\,km, due to their lower values for $\Lambda_{\rm s}$. The most direct comparison for our radius is with the GW-inferred value; \citet{Abbott2018} find $R_{\rm GW}=10.8^{+2.0}_{-1.7}$\,km (without imposing any EoS constraints \textit{a priori}). Our measurement is in excellent agreement with this value, and has a 90\% credible bound that is smaller by $\sim0.5$\,km. 

\citet{Abbott2018} find a larger value for the radius and a tighter posterior distribution, $R_{\rm GW}=11.9\pm1.4$\,km, if they impose a constraint that the radius must be compatible with equations of state that support a TOV mass larger than the most massive known NS. In a similar vein, we can use our constraint on \Mtov\ to tighten our posteriors on the NS radii. This assumes that the posteriors of $\tilde{\Lambda}$ and \Mtov\ are uncorrelated (Figure \ref{fig:cornershock} shows this to be a good approximation). We download the mass-radius curves for a sample of 4000 equations of state from \citet{Dietrich2020}, constructed using chiral effective theory (up to $1.5$ times nuclear saturation density, and a speed-of-sound parameterisation at higher densities) -- an approach that is motivated by nuclear theory \citep{Tews2018,Capano2020}.
As shown in the bottom panel of Figure \ref{fig:radius}, we select only those that support a TOV mass within our 90\% confidence interval. The inset histogram shows the distribution of \Rns\ for this EoS subset. We take this as the \textit{a priori} probability of a given radius in order to re-weight our posteriors for $R_1$ and $R_2$ \citep[after first transforming to a prior flat in \Rns\ following][]{Raithel2020}, resulting in the middle panel of Figure \ref{fig:radius}. Our final measurement using $\Lambda_{\rm s}$ and ensuring consistency with \Mtov\ is $R_1=11.06^{+1.01}_{-0.98}$\,km.

Similar to \citet{Abbott2018}, imposing a constraint on the EoS results in an increase in \Rns\ and a tightening of the posterior. Our final value is consistent with theirs at around the $1\sigma$ level, and narrower by $\sim 0.8$\,km. We also compare to multimessenger constraints from the literature that used the kilonova data for GW170817. Our median value for the radius lies just below the [11.3,13.5]\,km credible region obtained by \citet{Coughlin2019}, though there is significant overlap between their bounds and our 90\% confidence interval. Our measurement is in excellent agreement with the $11.0^{+0.9}_{-0.6}$\,km found by \cite{Capano2020}.
It is also consistent with the $11.75^{+0.86}_{-0.81}$\,km found by \citet{Dietrich2020}, at around the $1\sigma$ level. \citet{Breschi2021} and \citet{Most2018} report slightly larger values for \Rns, finding it to be $12.16^{+0.89}_{-1.11}$\,km or $12.39^{+1.06}_{-0.39}$\,km, respectively, which are consistent with our results only at the $\sim 2\sigma$ level.

\begin{figure*}
    \centering
    \includegraphics[width=\textwidth]{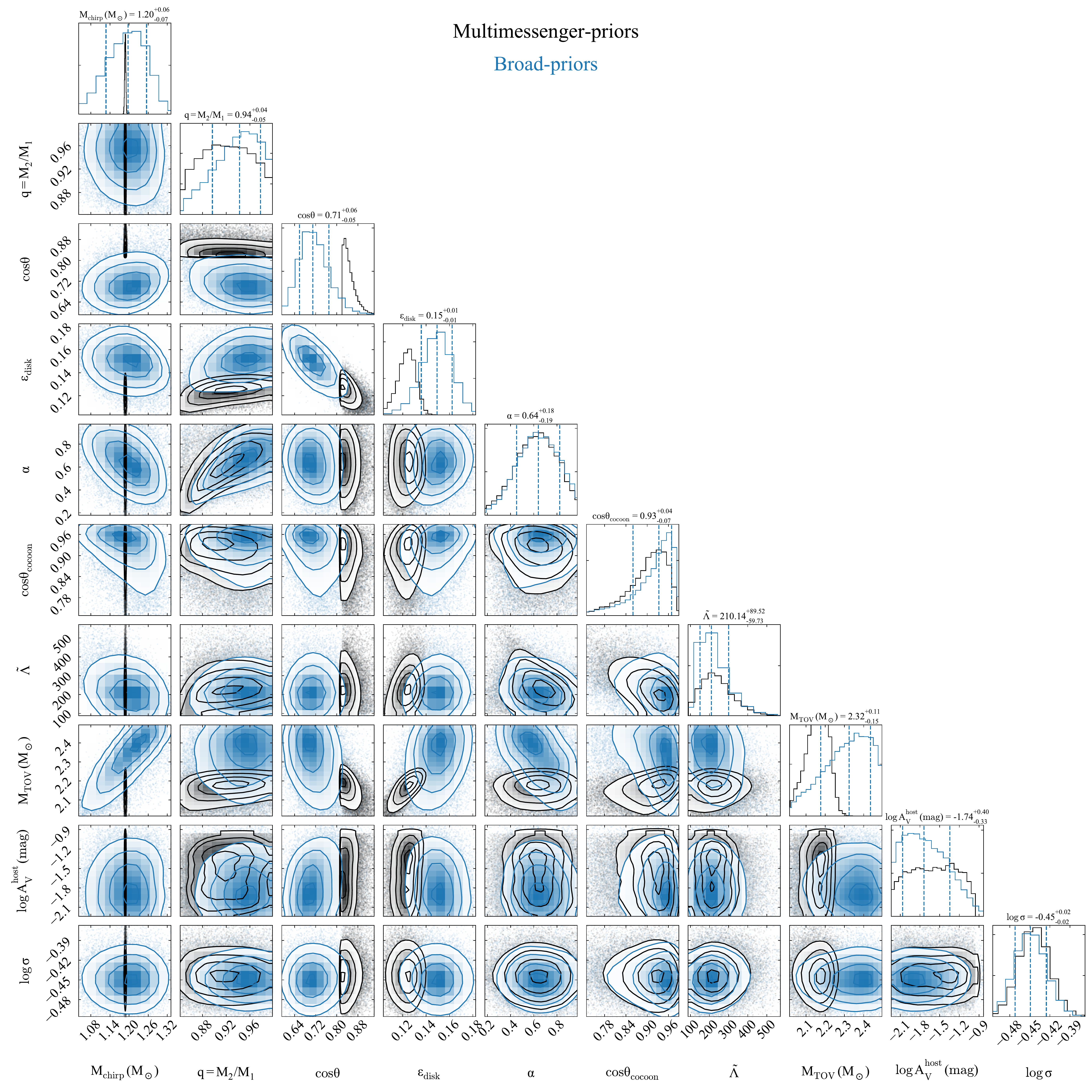}
    \caption{Posterior distributions for model fit to GW170817 when using the broad priors (blue), compared to the posteriors obtained using the multi-messenger priors (black; same as Figure \ref{fig:cornershock}). Both sets of priors are listed in Table \ref{tab:params}.}
    \label{fig:cornerbroad}
\end{figure*}

\subsection{The importance of multi-messenger data}

In this section we emphasise the importance of the \emph{multi-messenger} constraints we used to fit GW170817. Table \ref{tab:params} lists two sets of priors: one set of suggested default priors for fitting an arbitrary NS merger, and one specifically tuned for GW170817. This latter set of priors, which we employed in the previous section, makes use of the GW constraints on $\mathcal{M}$, $q$, $t_0$ and $\tilde{\Lambda}$, GRB afterglow constraints on $\theta$ (the GRB also constrains $t_0$), and constraints on extinction from analysis of the host galaxy \citep{Blanchard2017,Levan2017,Pan2017}.

Taking our preferred model for GW170817 (the agnostic model including both shock cooling and an enhanced surface wind ejecta), we change the priors to the default set and re-run the fit to GW170817. Figure \ref{fig:cornerbroad} shows the difference in posteriors when using the default priors, compared to the results using the multi-messenger priors. We find a fit of similar quality ($\ln{Z}=79.2$), but with much broader posterior distributions for several parameters.

The posterior distribution of $\mathcal{M}$ obtained using the broad priors has a median $\mathcal{M}_{\rm broad}=1.2$\,\M, close to the known value from the GW data. However, the credible region is $\sim 30$ times wider than when using the very tight GW constraints. Another notable difference is in the viewing angle, where much larger viewing angles $\theta_{\rm broad}\sim45^\circ$ are preferred without the GRB constraints. The distribution of mass ratio spans a similar range as in the multi-messenger case, but is skewed towards more equal systems ($q_{\rm broad}\sim0.94$). 

The posteriors of parameters that use the same priors in the default and multi-messenger models are (unsurprisingly) affected less severely, e.g.~the posteriors of $\alpha$ and $\cos{\theta_{\rm c}}$ are very similar in the two cases, as are the posteriors of $\tilde{\Lambda}$. However, the weaker constraints on the total system mass lead to a broadening of the posterior distribution of \Mtov, and a skew towards larger values.

The takeaway message is that our model can do a reasonable job of measuring important system properties for a GW170817-like merger even without a GW detection. However, having information from electromagnetic, GW, GRB and host galaxy analyses leads to tighter constraints, and this effect is perhaps most important when attempting to probe the NS EoS, as in section \ref{sec:eos}.

\begin{figure}
    \centering
    \includegraphics[width=\columnwidth]{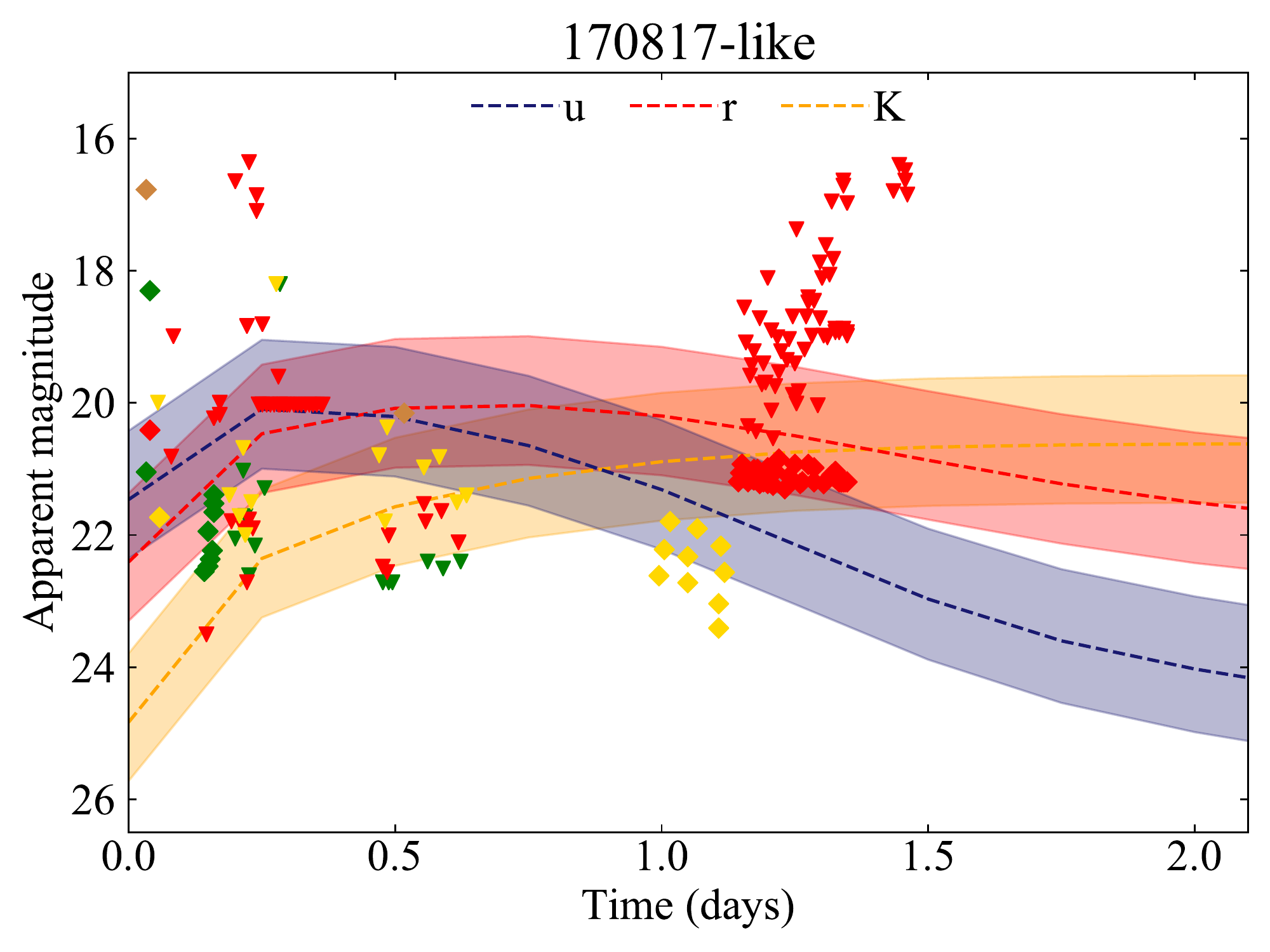}
    \includegraphics[width=\columnwidth]{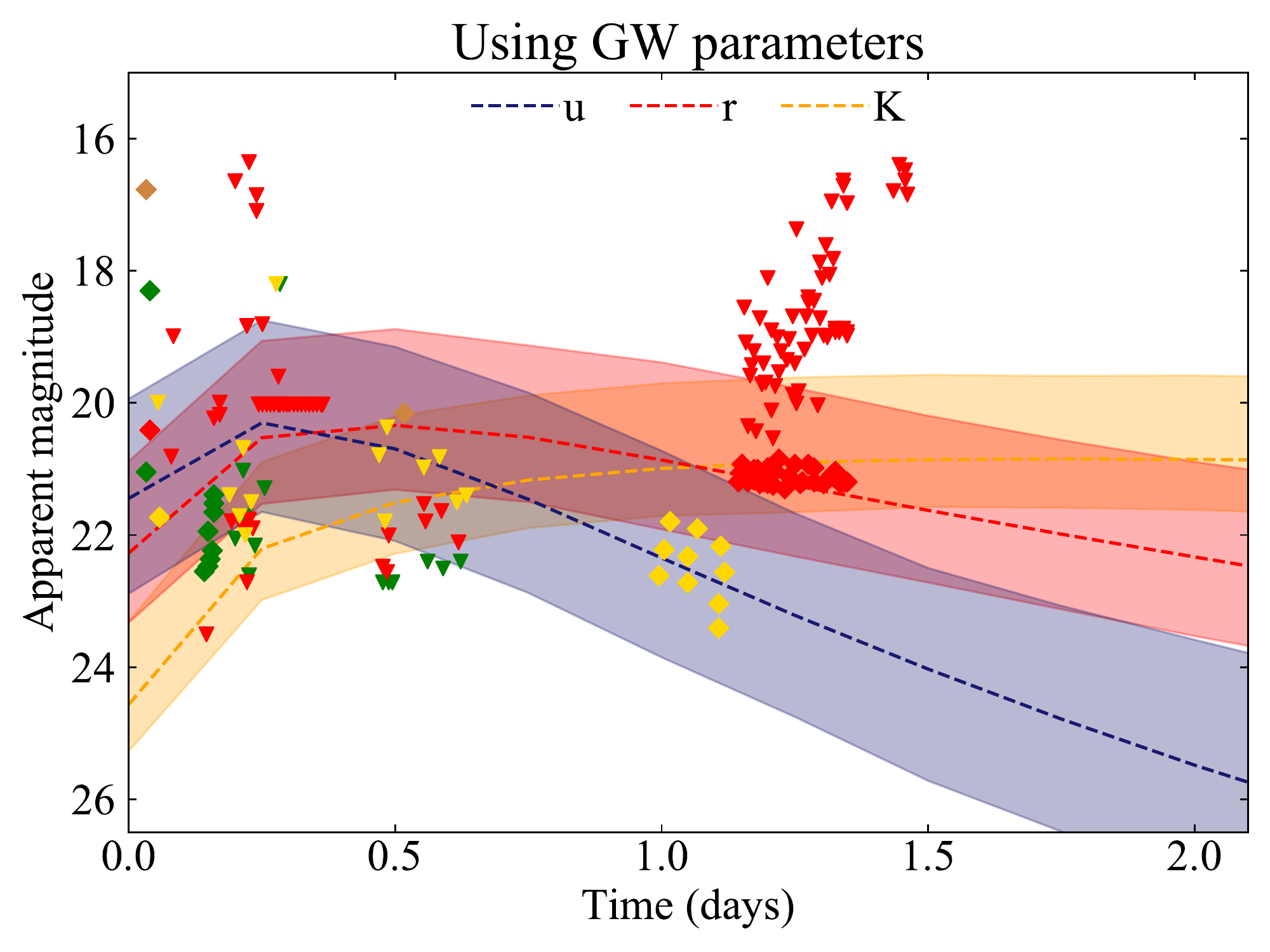}
    \caption{Model light curves for a GW190425 kilonova compared to limits compiled by \citet{Hosseinzadeh2019}. The shaded regions correspond to the 90\% credible ranges in magnitude for each band. Top: assuming a GW170817-like kilonova shifted to the GW-inferred distance of GW190425 (following \citealt{Hosseinzadeh2019}). Bottom: predicting the light curve directly from the GW constraints on the chirp mass, mass ratio, distance and inclination. The wider uncertainty bands and faster decline in this case show that assuming a GW170817-like event leads to unrealistically tight constraints and observations that may be too shallow to confidently detect this source.}
    \label{fig:190425}
\end{figure}

\section{Light curve predictions: example of GW190425}
\label{sec:predict}

Having validated our model using GW170817, we now (briefly) demonstrate its utility for generating simulated kilonova data. While the broader applications in terms of population synthesis will be explored in future works, an important use case that we discuss here is to predict the electromagnetic observables of a particular kilonova given a detected GW signal (see e.g. \citealt{Margalit2019}), in order to make follow-up observations more efficient. This is especially relevant given the possibility of target-of-opportunity GW follow-up with the upcoming Vera Rubin Observatory \citep{Margutti2018b,Cowperthwaite2019,Smith2019,Chen2020}.

During the LIGO-Virgo O3 run, the GW detectors identified several binary NS candidates. One event is contained in the latest GW source catalog, GWTC-2 \citep{Abbott2020c}, with the others discovered in the second half of O3 such that source parameters have not yet been released. The published event is GW190425, which was notable for its large total mass $M_1+M_2=3.4$\,\M\ \citep{Abbott2020a}, implying that (unlike GW170817) at least one of the constituent NSs was more massive than the canonical $\approx 1.4$\,\M. 

Various groups of astronomers followed up the GW discovery of GW190425 with electromagnetic observations in an attempt to detect a counterpart, though without success on this occasion \citep{Hosseinzadeh2019,Coughlin2019b,Lundquist2019,Antier2020}. This may have been due in part to the large distance to the source, $159^{+69}_{-71}$\,Mpc, and in part because the signal was significant only in one GW detector, leading to a wide sky localization of $>8000\,{\rm deg}^2$ \citep{Abbott2020a}.

\citet{Hosseinzadeh2019} compiled detection limits from both galaxy-targeted and publicly-reported wide-field searches, and compared to the light curves of GW170817 (shifted to the distance of GW190425). In Figure \ref{fig:190425}, we show these limits compared to our best-fit model of GW170817 in the left panel. In agreement with \citet{Hosseinzadeh2019}, we see that the deep galaxy-targeted searches rule out a GW170817-like counterpart (in those particular galaxies) over most of the 90\% plausible distance range. 

With our new model, we can use the chirp mass and mass ratio constraints obtained by LIGO and Virgo for GW190425 to predict a light curve more directly applicable to this source. A similar approach was adopted by \citet{Barbieri2020} for both NS-NS and NS-BH binaries. This prediction is not only tuned specifically to GW190425, but also folds in the uncertainties on the binary parameters as well as the distance uncertainty. We use a Gaussian distribution for the chirp mass $\mathcal{M}=1.44\pm0.013$, and flat distributions $0.8<q<1$, $0<\cos{\theta}<1$, $0.71<\cos{\theta_{\rm c}}<1$, and a fiducial $\epsilon_{\rm disk}=0.15$ based on GW170817. The result is shown in the right panel of Figure \ref{fig:190425}.

Two features of this prediction are noteworthy. One is the increased width of the 90\% distribution of credible light curves, allowing more space for a faint kilonova to go undetected at the observed depths. More significantly, the median light curve is fainter (by $\sim 0.7$\,mag in $r$ band at 1 day post-merger) and declines more rapidly than GW170817. The main reason for this is that the more massive remnant in this system is expected to collapse promptly to a BH, reducing the mass of the purple ejecta component by an order of magnitude compared to GW170817. With this faster decline, even many of the galaxy-targeted observations obtained $\gtrsim1$\,day after merger rule out only $\sim 50\%$ of the plausible light curves.

Our goal here is not to calculate precisely the fraction of the localization volume probed by optical observations of GW190425, but simply to make the general point that how constraining a detection limit is depends on the choice of comparison model, and that this in turn depends on the binary source parameters. A better strategy in future would be to tune the observed depths to a level necessary to cover the 90\% credible range of kilonova luminosities \emph{for a given GW detection}. At present, this is not possible, as the GW-inferred binary parameters needed to simulate the light curve are not released at the time of merger. We suggest that greater cooperation between the GW and electromagnetic astronomy communities could be very useful in this regard: if the GW source parameters ($\mathcal{M}$, $q$, $\tilde{\Lambda}$) were released in low latency, we could use our model to immediately predict the required observing depth at any wavelength to find or rule out a kilonova counterpart at high significance.

\section{Conclusions}
\label{sec:conc}

We have presented a simple forward model that combines a range of literature results on ejecta masses, r-process opacities, and the NS EoS, to predict kilonova light curves directly from GW-accessible binary parameters. This model can also be used to derive limits on the kilonova rate from optical surveys, given a model for the BNS source population; this will be the subject of future work. Our code has been developed within the \textsc{mosfit} framework and is publicly available.

We validated the model by fitting it to the well-observed kilonova associated with GW170817, incorporating GW and GRB afterglow results in our priors. We found that a model including a shock cooling cocoon component was overwhelmingly favoured by the Bayes factor, compared to models with r-process radioactive decay as the only luminosity source. Data obtained during the first day after merger was essential to differentiating these models \citep{Arcavi2018}. Our approach directly associates each of the r-process ejecta components with a physical origin, where the low opacity material is shock-heated dynamical ejecta, and the higher opacity material that dominates the emission from $\sim 2-10$ days after merger is a viscous disk wind. The very lanthanide-rich tidal ejecta makes a significant contribution only at very late times.

The posteriors of our model fit contain important information on the GW170817 progenitor system. We find that this system likely had a small mass ratio, $q\approx0.9$, and was viewed $\approx 30^\circ$ off-axis. By including the maximum stable NS mass \Mtov\ and the symmetric tidal deformability $\Lambda_{\rm s}$ in our free parameters, we constrained these EoS-dependent parameters. We measure \Mtov\,$=2.17^{+0.08}_{-0.11}$\,\M\ (90\% credible interval) and derive \Rns\,=$11.06^{+1.01}_{-0.98}$\,km using the posteriors of $\Lambda$, $\mathcal{M}$, $q$ and \Mtov, carefully taking into account the systematic errors on all relations used. This radius constraint is consistent and competitive with others in the literature.

For high-significance detections in O3, the LIGO-Virgo collaboration provided publicly the distance and source classification probabilities (as NS, BH, NS-BH, or mass-gap binary) along with the sky localisation, giving astronomers a rough guide to sources they may wish to follow up electromagnetically. However, even for a source confidently classified as a NS merger, the binary parameters -- especially the chirp mass, mass ratio and viewing angle -- have a substantial impact on the likelihood of detecting the counterpart, since they control the peak luminosity, colour and decline rate (Figures \ref{fig:mass} and \ref{fig:examples}). If such parameters (which may be estimated even in low latency; \citealt{Biscoveanu2019,Finstad2020,Krastev2020}) are made available to the astronomical community, this can aid real-time kilonova searches following binary NS merger discoveries from the GW detector network \citep[also advocated for by][for example]{Margalit2019,Barbieri2020}. We demonstrated this by synthesising light curves appropriate to GW190425, for which optical searches did not uncover a counterpart, finding that many were not deep enough to fully rule out a kilonova. For future observing runs with a higher binary NS detection rate, the ability to better predict the required observing depth on a case-by-case basis would enable more efficient allocation of telescope time (which is of course a finite resource) for GW follow-up.

We have not discussed in this work the signature of a merger between a NS and a companion BH; these are also expected to produce kilonovae as long as the mass ratio is modest (though without shock-driven, low-lanthanide ejecta). Our next step will be to implement an analogous model to simulate light curves for this class of events \citep[see also][]{Barbieri2020}. 

Finding a larger sample of kilonovae in the coming years, as GW detectors and optical telescopes increase their survey power, will be crucial to improve our understanding of the NS EoS and the source population of NS binaries. Hopefully, the code we present here will be useful to the community both in helping to plan observations that increase this sample and in deriving important physics from the data obtained.

\section*{Acknowledgements}

We thank the referee Rahul Kashyap for valuable suggestions that improved this paper, Jenni Barnes for providing kilonova thermalisation data, and 
Edo Berger, Ryan Chornock, Laurence Datrier, Michael Fulton, Martin Hendry, Albert Kong, Gavin Lamb, Brian Metzger, Christopher Moore, Geraint Pratten, Surojit Saha, Stephen Smartt, Alberto Vecchio and Ashley Villar for their insights and discussions.
MN is supported by a Royal Astronomical Society Research Fellowship. This project has received funding from the European Research Council (ERC) under the European Union’s Horizon 2020 research and innovation programme (grant agreement No.~948381).
BM is supported by NASA through the NASA Hubble Fellowship grant \#HST-HF2-51412.001-A awarded by the Space Telescope Science Institute, which is operated by the Association of Universities for Research in Astronomy, Inc., for NASA, under contract NAS5-26555.
PS is supported by the Dutch Research Council (NWO) Veni Grant No. 680-47-460.
GPS acknowledges support from the Science and Technology Facilities Council through grant number ST/N021702/1.

\section*{Data Availability}

 This paper makes use of existing public data from \url{kilonova.space}. The code has been made available at \url{https://github.com/guillochon/MOSFiT}. A population of kilonova models produced by this code using broad priors, in UV, optical and NIR filters, are available for download at \url{https://github.com/mnicholl/kn-models-nicholl2021}.



\bibliographystyle{mnras}
\bibliography{refs} 




\appendix

\section{New fits to simulation data}

\begin{figure*}
    \centering
    \includegraphics[width=\columnwidth]{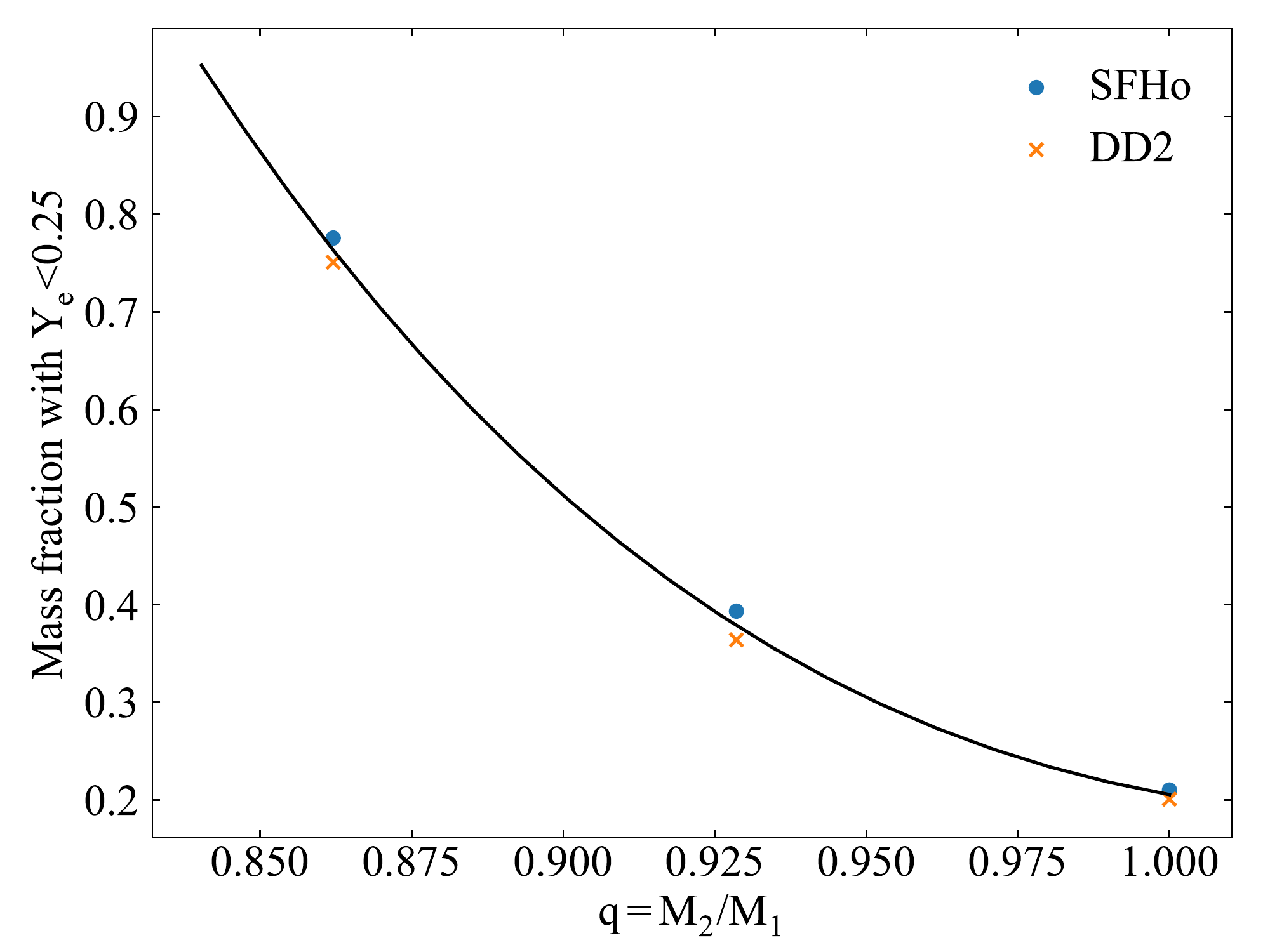}
    \caption{Mass fraction of lanthanide-poor dynamical ejecta as a function of the binary mass ratio, $q=M_2/M_1$, using $Y_e$ distributions from \citet{Sekiguchi2016} for the SFHo (soft) and DD2 (stiff) equations of state. Our polynomial fit predicts essentially no lanthanide-poor ejecta for $q\lesssim 0.8$, in agreement with \citet{Dietrich2017}.}
    \label{fig:ye}
\end{figure*}

Figure \ref{fig:ye} shows our quadratic fit to determine $f_{\rm red}$ and $f_{\rm blue}\equiv 1-f_{\rm red}$, the fractions of red (high opacity) and blue (low opacity) dynamical ejecta, using the simulated $Y_e$ distributions from \citet{Sekiguchi2016}. Ejecta with $Y_e\le 0.25$ are assumed to be red.

\begin{figure*}
    \centering
    \includegraphics[width=\columnwidth]{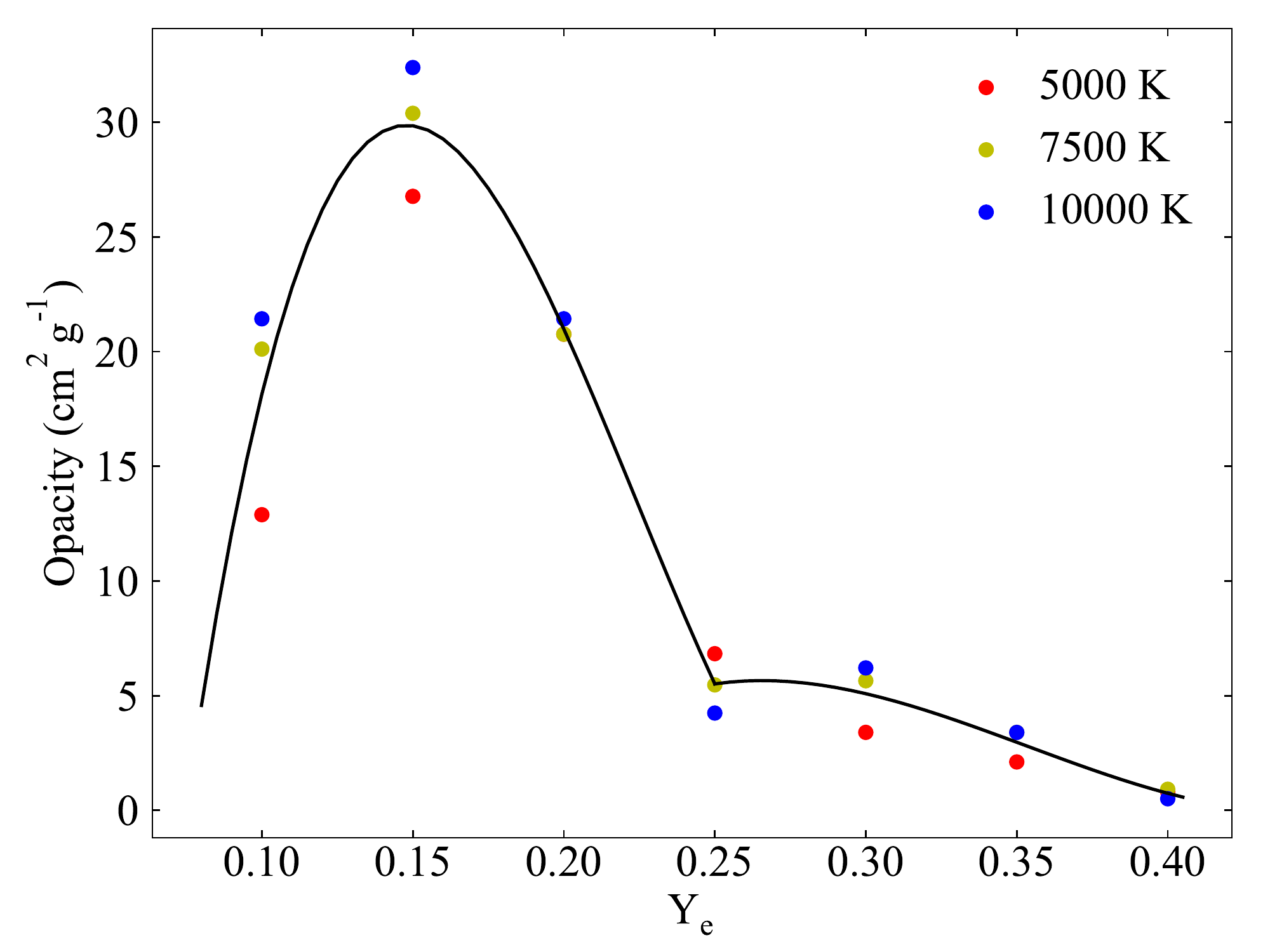}
    \caption{Opacity of r-process ejecta as a function of electron fraction and temperature from \citet{Tanaka2020}. The broken polynomial fit is used to estimate the opacity of the disk ejecta, using the average $Y_e$ predicted by \citet{Lippuner2017} for stable, short-lived or prompt-collapse merger remnants. In practice, the average $Y_e>0.25$ for all reasonable input parameters.}
    \label{fig:kappa}
\end{figure*}

Figure \ref{fig:kappa} shows our broken polynomial fit to the simulations of \citet{Tanaka2020}, which we use to estimate the opacity of the purple (post-merger) ejecta as a function of $Y_e$.

\section{Plots of additional light curve fits}

\begin{figure*}
    \centering
    \includegraphics[width=\columnwidth]{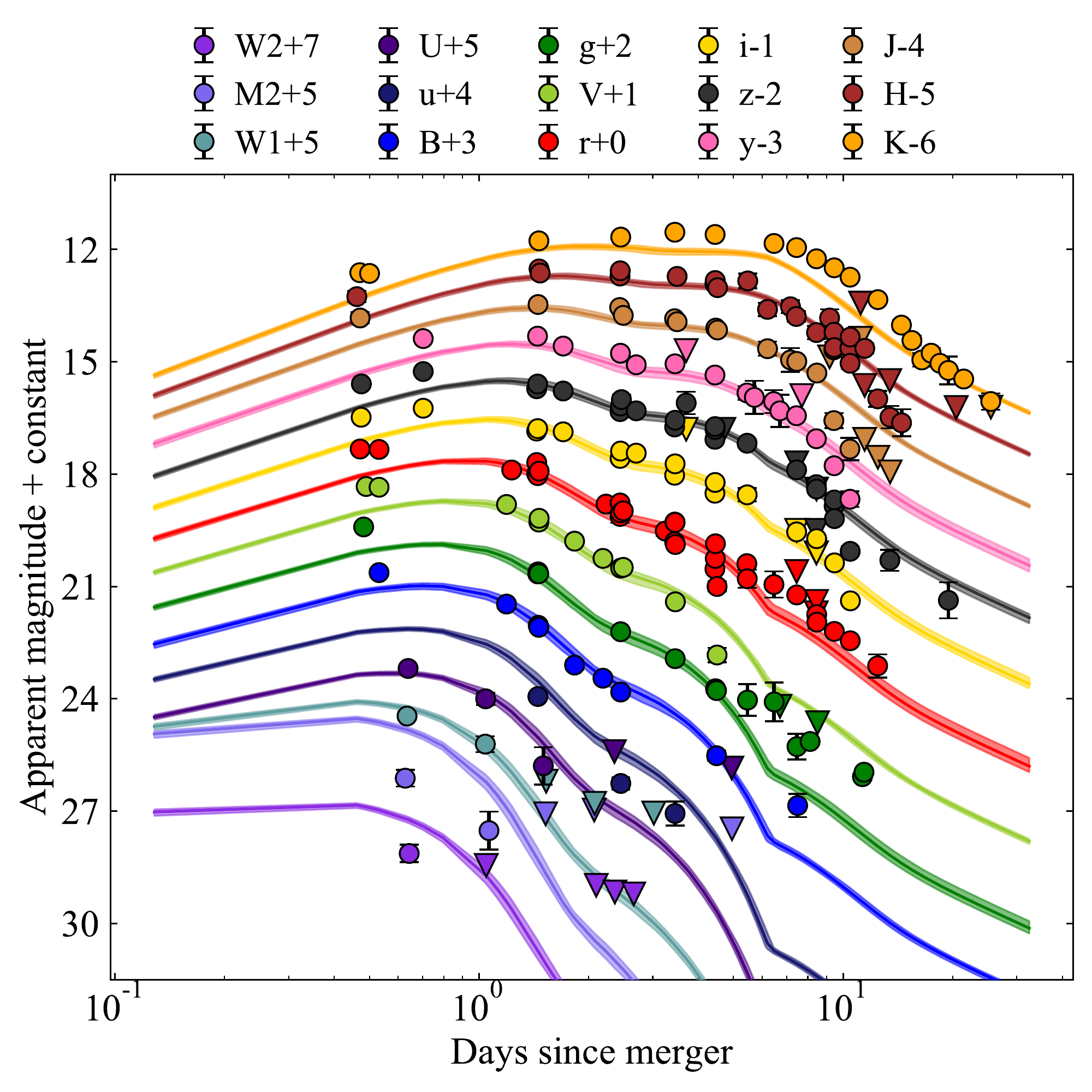}
    \includegraphics[width=\columnwidth]{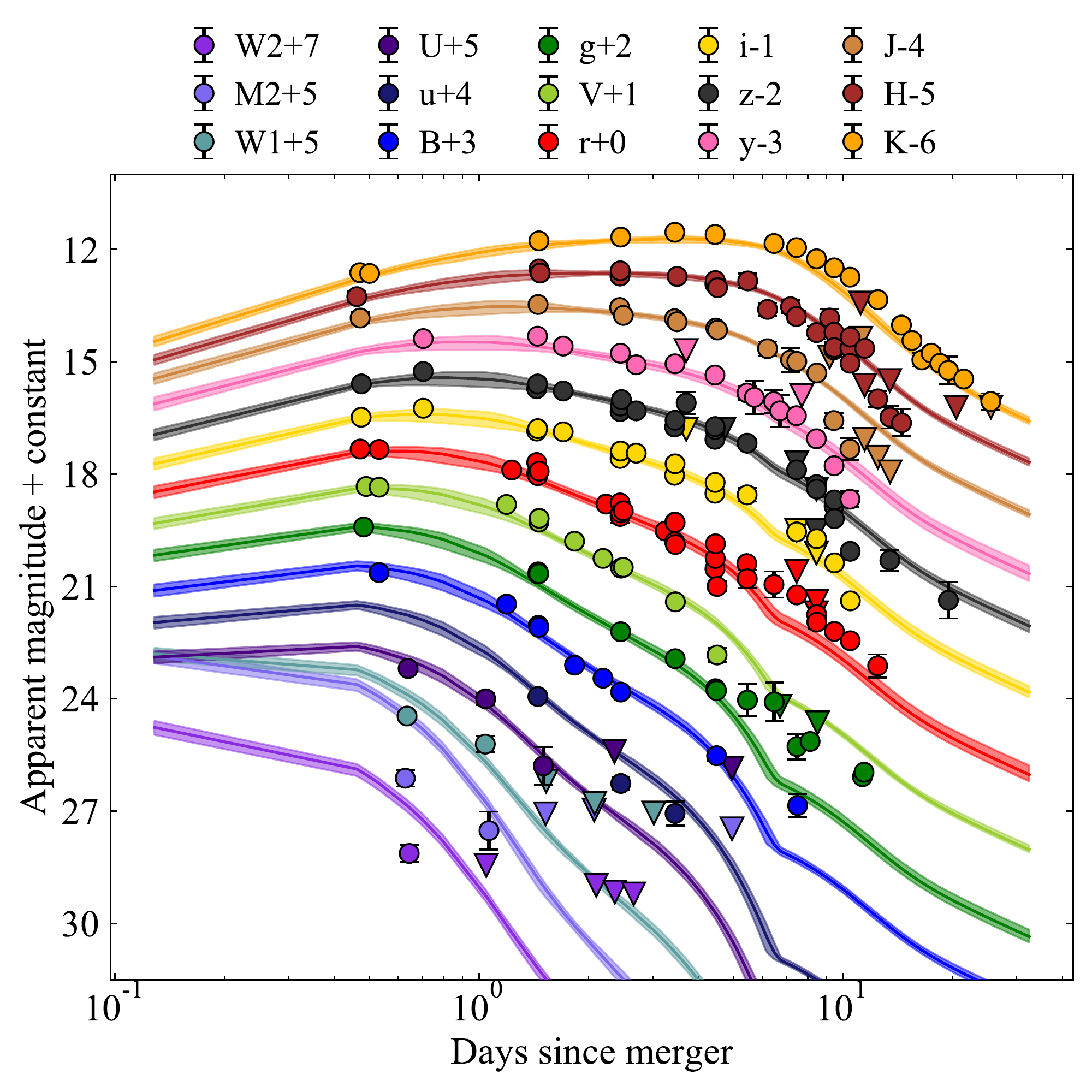}
    \caption{Left: Best-fit kilonova model for GW170817 with surface ejecta enhancement. Right: Best-fit model with shock cooling. The shock cooling model is preferred with a Bayes factor $B>10^{10}$. This figure is similar to Figure \ref{fig:lc} in the main text, which shows the best fit light curves including either both or neither of these effects.}
    \label{fig:morelcs}
\end{figure*}

Figure \ref{fig:morelcs} shows the light curve fits for the surface-wind-only and shock-only versions of the model.

\bsp	
\label{lastpage}
\end{document}